\documentclass[sigconf]{acmart}

\usepackage[yyyymmdd]{datetime}

\usepackage{tabularx}
\usepackage{dcolumn} 
\newcolumntype{d}[1]{D{.}{.}{#1}}

\usepackage{subcaption}

\usepackage{booktabs}
\usepackage{multirow}
\usepackage{array}
\usepackage{tikz}
\usetikzlibrary{shapes}
\usepackage{todonotes}

\let\xtodo\todo
\renewcommand{\todo}[1]{\xtodo[inline,color=green!50]{#1}}

\usepackage[strings]{underscore}
\usepackage{colortbl}
\usepackage{xcolor}
\usepackage{hyperref}
\AtBeginDocument{%
  }

\copyrightyear{2026}
\acmYear{2026}
\setcopyright{cc}
\setcctype{by}
\acmConference[CHI '26]{Proceedings of the 2026 CHI Conference on Human Factors in Computing Systems}{April 13--17, 2026}{Barcelona, Spain}
\acmBooktitle{Proceedings of the 2026 CHI Conference on Human Factors in Computing Systems (CHI '26), April 13--17, 2026, Barcelona, Spain}
\acmPrice{}
\acmDOI{10.1145/3772318.3791191}
\acmISBN{979-8-4007-2278-3/2026/04}

\begin{document}

\title[Sensing What Surveys Miss]{Sensing What Surveys Miss: Understanding and Personalizing Proactive LLM Support by User Modeling}


\settopmatter{authorsperrow=2}

\author{Ailin Liu}
\orcid{0009-0002-7489-058X}
\affiliation{%
  \institution{LMU Munich}
  \institution{Munich Center for Machine Learning (MCML)}
  \country{Germany}}
\email{ailin.liu@lmu.de}

\author{Yesmine Karoui}
\orcid{0009-0003-9027-5795}
\affiliation{
  \institution{LMU Munich}
  \country{Germany}
}
\email{y.karoui@campus.lmu.de}

\author{Fiona Draxler}
\orcid{0000-0002-3112-6015}
\affiliation{
  \institution{University of Mannheim}
  \country{Germany}
}
\email{fiona.draxler@uni-mannheim.de}

\author{Frauke Kreuter}
\affiliation{
  \institution{LMU Munich}
  \institution{Munich Center for Machine Learning (MCML)}
  \country{Germany}
}
\email{frauke.kreuter@stat.uni-muenchen.de}

\author{Francesco Chiossi}
\orcid{0000-0003-2987-7634}
\affiliation{
\institution{LMU Munich}
\country{Germany}
}
\email{francesco.chiossi@um.ifi.lmu.de}

\renewcommand{\shortauthors}{Liu et al.}

\begin{abstract}

Difficulty spillover and suboptimal help-seeking challenge the sequential, knowledge-intensive nature of digital tasks. In online surveys, tough questions can drain mental energy and hurt performance on later questions, while users often fail to recognize when they need assistance or may satisfy, lacking motivation to seek help. We developed a proactive, adaptive system using electrodermal activity and mouse movement to predict when respondents need support. Personalized classifiers with a rule-based threshold adaptation trigger timely LLM-based clarifications and explanations. In a within-subjects study (N=32), aligned-adaptive timing was compared to misaligned-adaptive and random-adaptive controls. Aligned-adaptive assistance improved response accuracy by 21\%, reduced false negative rates from 50.9\% to 22.9\%, and improved perceived efficiency, dependability, and benevolence. Properly timed interventions prevent cascades of degraded responses, showing that aligning support with cognitive states improves both the outcomes and the user experience. This enables more effective, personalized LLM-assisted support in survey-based research.

\end{abstract}

\ccsdesc[300]{Human-centered computing~Web-based interaction}
\ccsdesc[300]{Human-centered computing~Ubiquitous and mobile computing systems and tools}
\ccsdesc[500]{Human-centered computing~Empirical studies in interaction design}
\ccsdesc[500]{Human-centered computing~Interactive systems and tools}

\keywords{Proactive Assistance, Physiological Sensing, Personalization, User Experience, Realtime Support}



\maketitle

\begin{figure*}[ht]
  \centering
  \begin{tabular}{ccc}
    \includegraphics[width=0.49\textwidth]{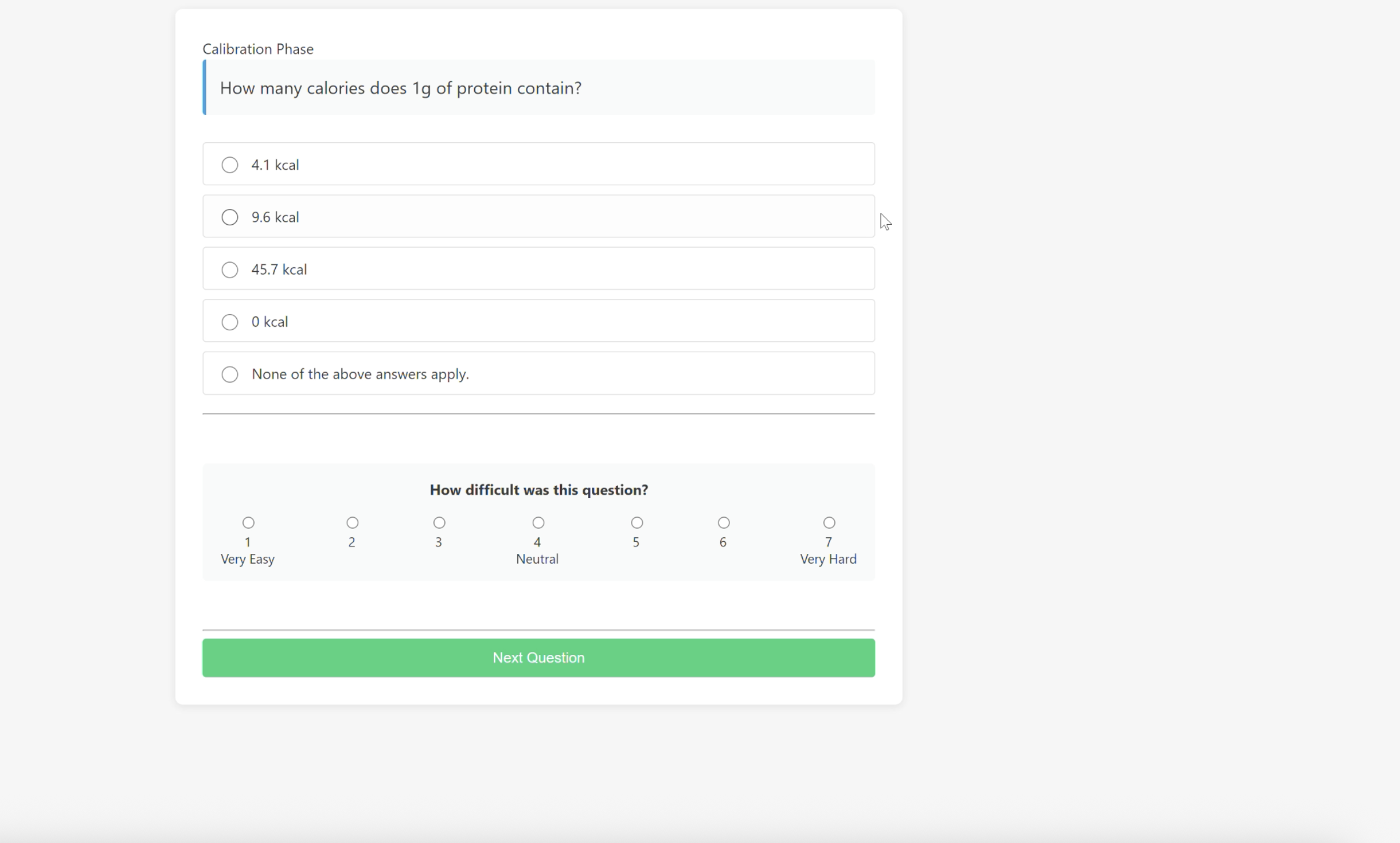} &
    \includegraphics[width=0.49\textwidth]{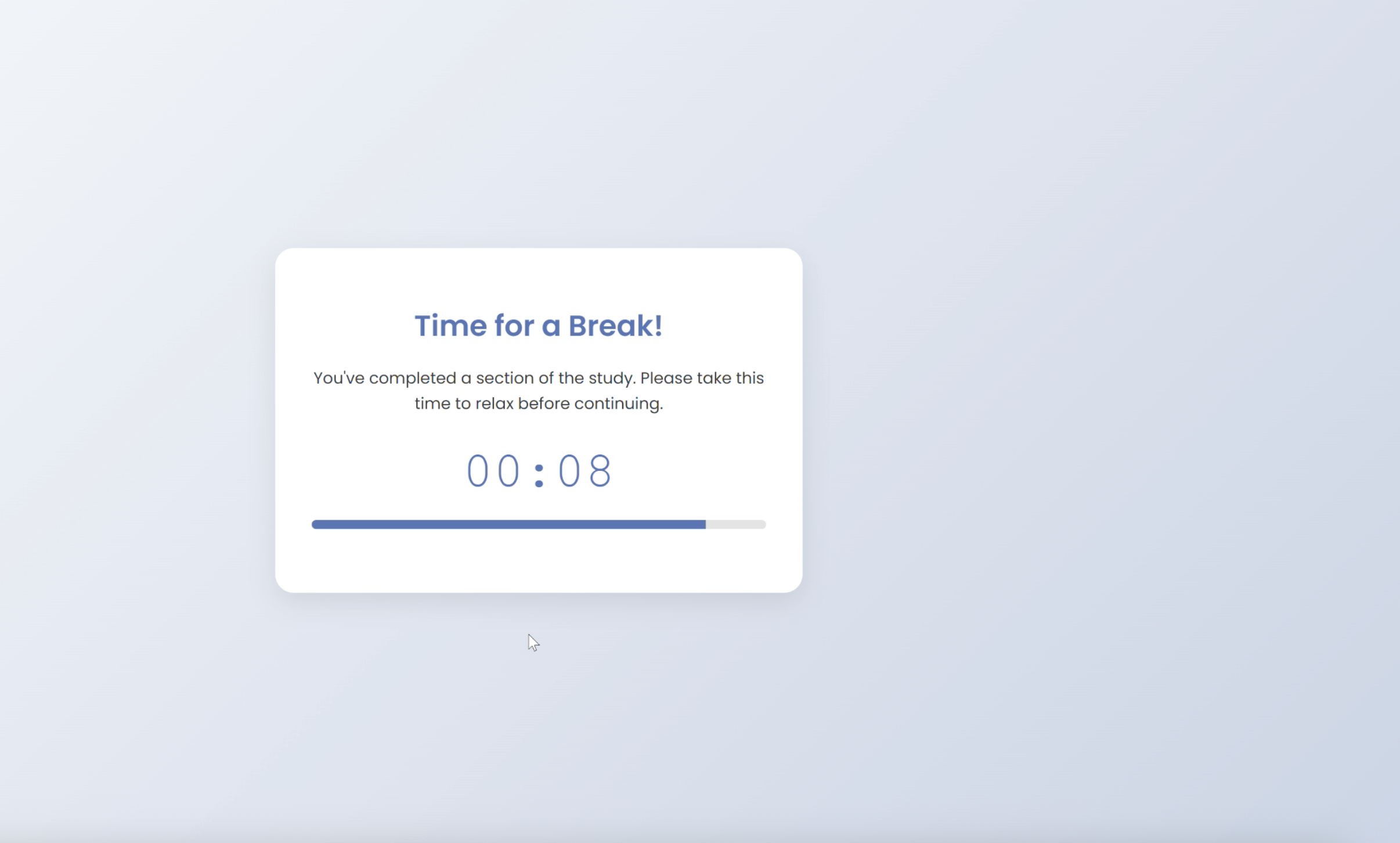} \\
    (a) Calibration Screen & (b) Rest Screen \\
    \includegraphics[width=0.49\textwidth]{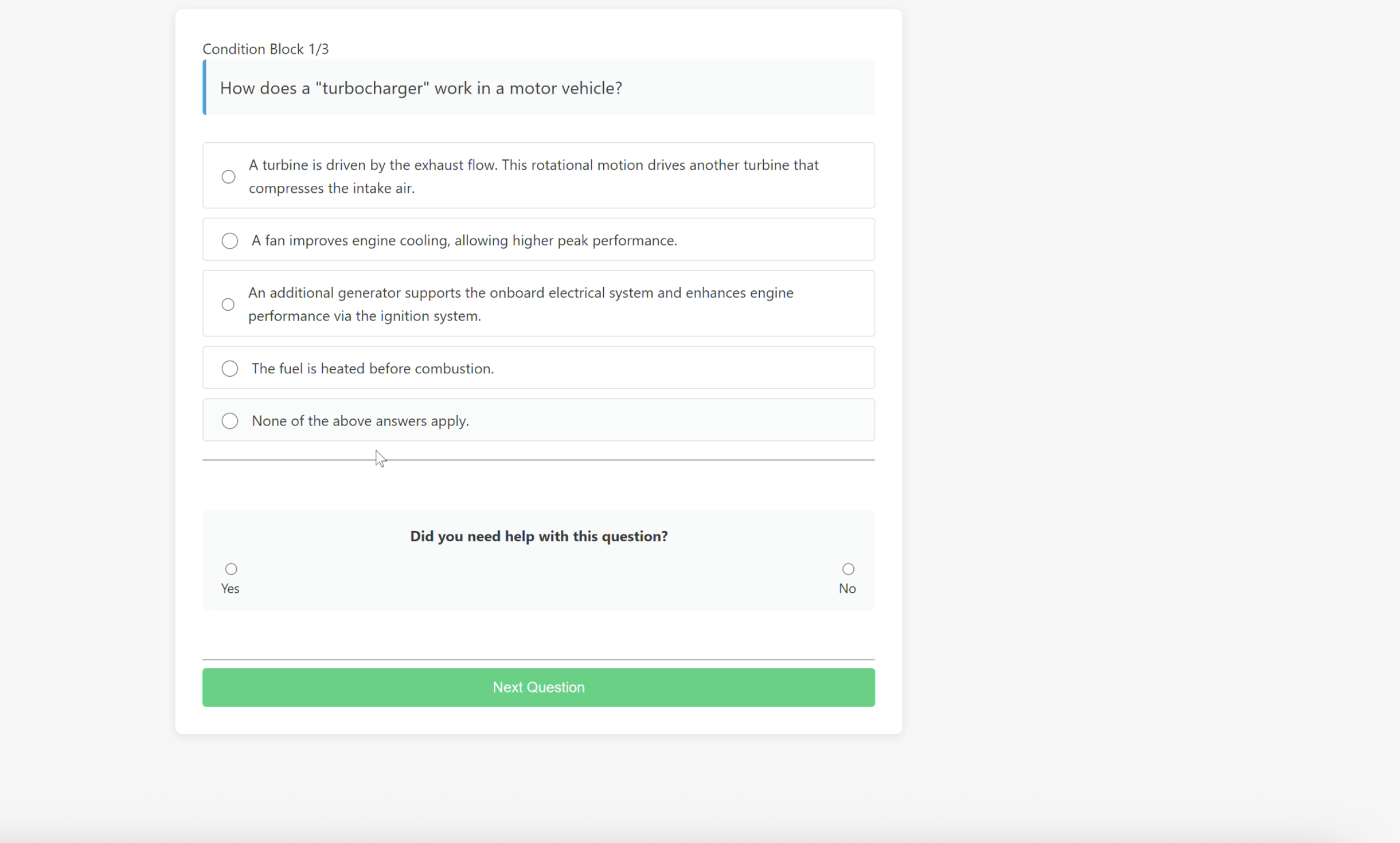} &
    \includegraphics[width=0.49\textwidth]{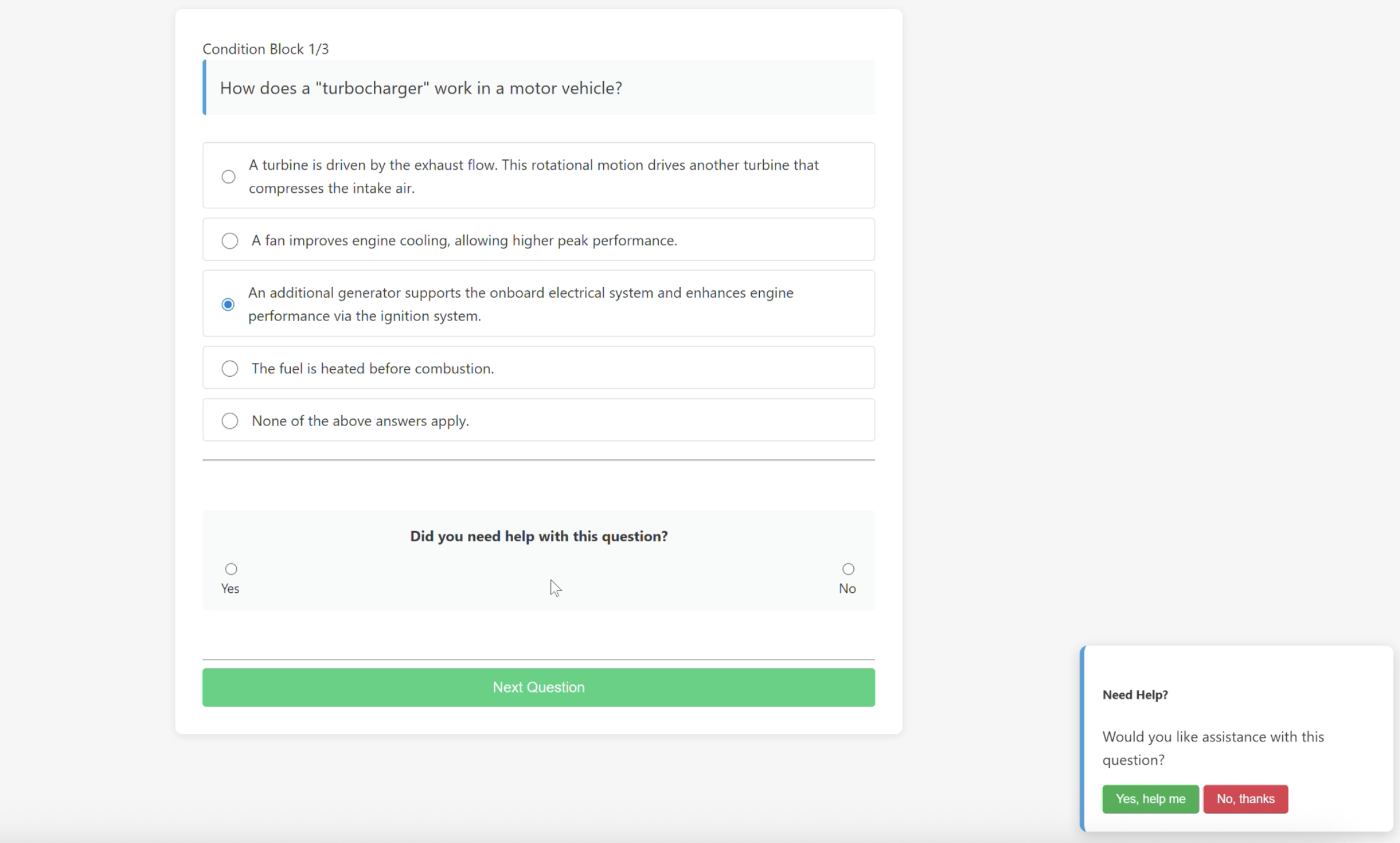} \\
    (c) Condition Screen & (d) Helper Triggered\\
    \includegraphics[width=0.49\textwidth]{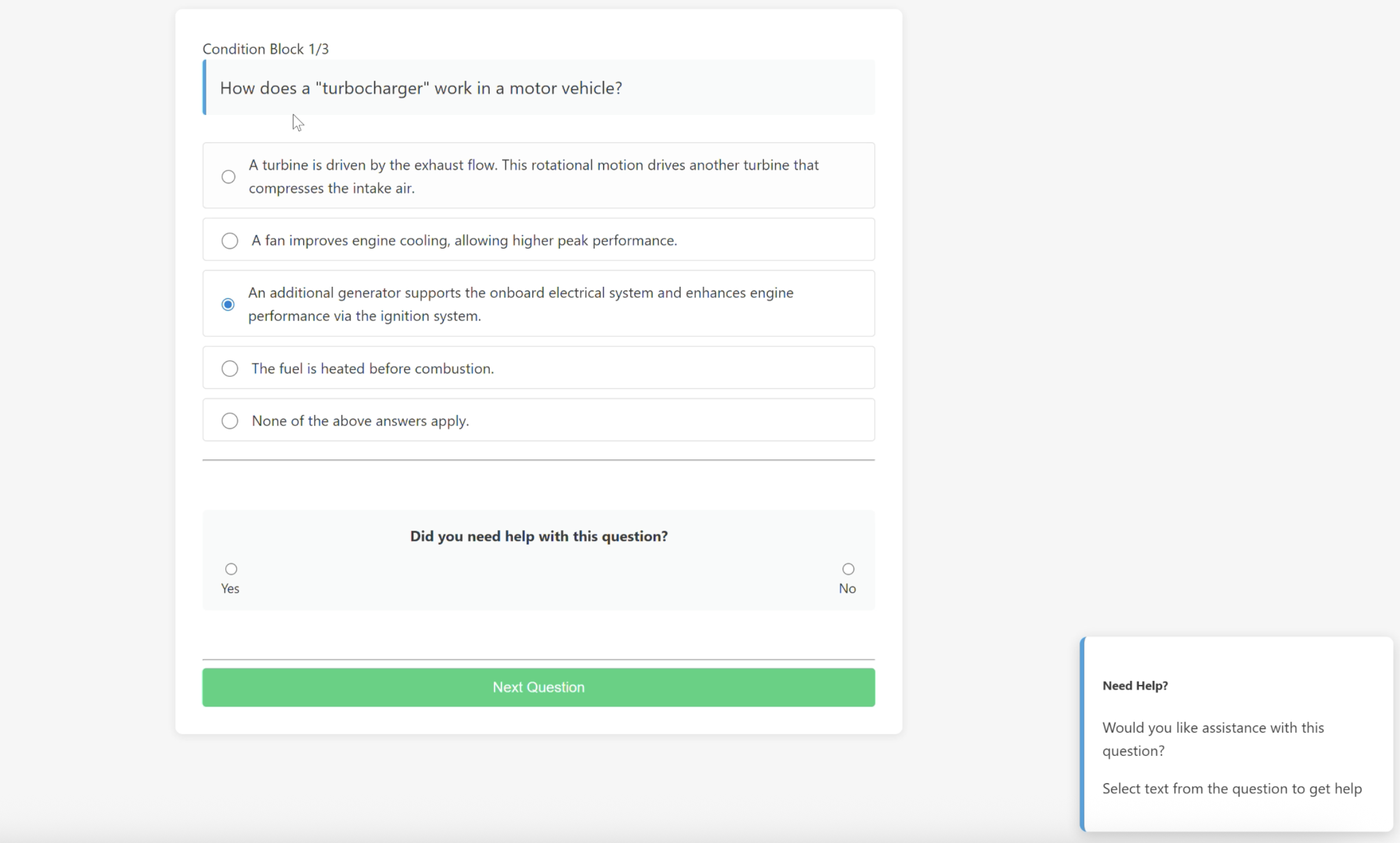} &
    \includegraphics[width=0.49\textwidth]{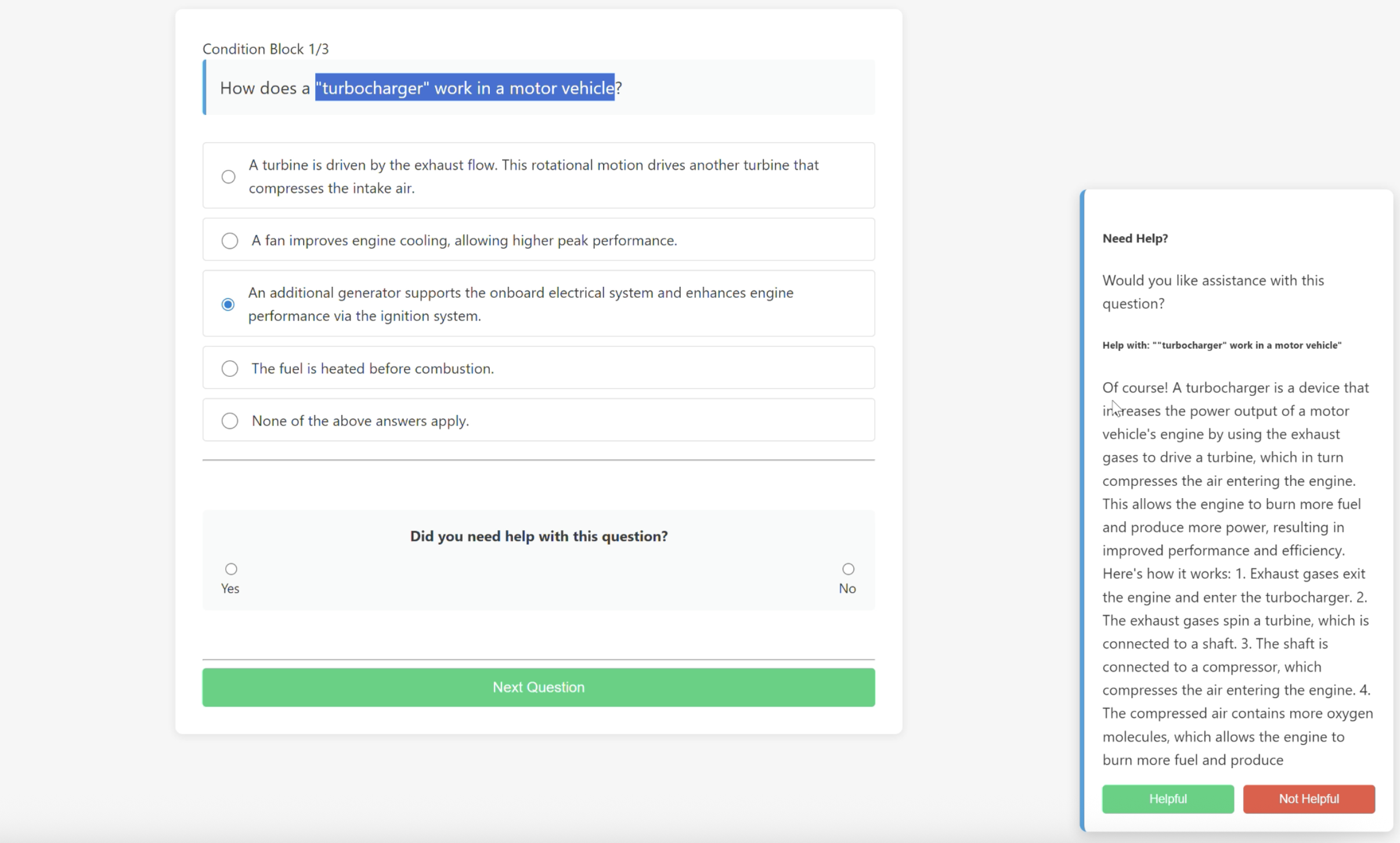} \\ 
    (e) Waited for Query Text & (f) Assistance Provided\\
  \end{tabular}
  \caption{Screenshots of the experimental interface: (a) calibration phase, (b) rest break, (c) condition task question, (d) adaptive helper triggered, (e) waiting for user selecting text, and (f) assistance provided by the system.}
  \label{fig:study-screenshots}
  \Description{Screenshots showing the system providing assistance to the user.}
\end{figure*}

\section{Introduction}
Surveys are a widely used method for collecting data across domains from product design and marketing to social research and academic feedback. Their ability to gather insights quickly, cost-effectively, and at scale makes them a core instrument for both researchers and practitioners. However, unlike interviewer-administered surveys, where trained professionals can detect hesitation, confusion, or disengagement and intervene in real time, self-administered web surveys provide no such adaptive human support. When respondents face cognitively demanding questions or feel pressured by time constraints, they may engage in satisficing behaviors, selecting suboptimal responses to reduce effort and ultimately compromising data quality \cite{Roberts2019, Blazek2023}. To mitigate these risks, survey methodologists have emphasized the importance of identifying problematic items and respondent difficulties, whether through pretesting before deployment, intervention during survey administration, or post-hoc detection of low-quality responses \cite{daikeler2024assessing, restuccia2017quality}.

This challenge is especially pronounced in surveys that include cognitive or general-knowledge items intended to assess participants’ understanding or abilities. Such questions are widely used in national social surveys, civic knowledge measures, and empirical research to measure comprehension \cite{Tourangeau2000, Frederick2005, gss2024, Cokely2012}. When respondents struggle with such questions, whether due to item complexity, ambiguous wording, or insufficient domain knowledge, the cognitive strain can have spillover effects on subsequent items, diminishing attentiveness and consistency \cite{ashley2023predicting}. While recent work suggests that the increase in perceived burden across items may be modest in some contexts \cite{kunz2025EffectsObjectivePerceived}, even small increases can compound in ways that influence overall engagement and accuracy. Over time, this strain can manifest as satisficing behaviors (e.g., straightlining, skipping items) \cite{Roberts2019, Blazek2023} or early termination, ultimately reducing both the completeness and validity of collected data. 

Survey methodology has investigated interactive support through requests for clarifications \cite{conrad2006use, schober2000clarifying, Conrad2000}. Meanwhile, large language models (LLMs) make it possible to deliver flexible, context-sensitive explanations in natural language \cite{Chen2025Need, imrie2023redefining, Buinca2025}.

However, existing approaches face fundamental limitations in their ability to 
provide timely, individualized support. Static personalization systems determine 
support strategies based on pre-task assessments (e.g., prior knowledge tests, 
demographic profiles) and apply the same intervention pattern to all items for 
a given user \cite{conrad2006use, schober2000clarifying, Conrad2000}. While these systems can tailor content to user characteristics, 
they cannot respond to within-session fluctuations in cognitive state, a critical 
gap given that respondent struggle varies moment-to-moment based on item difficulty, 
accumulated fatigue, and interference from prior items.

Reactive support systems wait for explicit user requests such as, help buttons and chat 
queries, before intervening \cite{conrad2003interactive, Conrad2007}. Previous research shows that respondents rarely request help on their own, often fewer than 10\% use optional clarifications and respondents had often already formed incorrect interpretations by the time the system reacted \cite{Conrad2007, conrad2006use, conrad2003interactive}.

Rule-based systems adjust based on fixed decision criteria (e.g., ``offer help 
after 30 seconds of inactivity''), but these universal thresholds ignore individual 
differences in baseline behavior—what signals struggle for one person may be 
normal processing for another \cite{horwitz2017using}. The core limitation across 
these approaches is their inability to detect and respond to the dynamic, 
personalized cognitive state of individual respondents in real time. When a 
respondent encounters an unexpectedly difficult item, existing systems either 
miss the moment of struggle entirely (static approaches), detect it too late 
(reactive approaches), or misfire by applying population-level heuristics 
(rule-based approaches). This timing gap is consequential: premature assistance 
interrupts productive problem-solving and reduces engagement, while delayed 
support arrives too late to prevent cascading effects on subsequent items 
\cite{Aleven2003, Roll2011}. What is needed are systems that continuously monitor 
individual cognitive load through leading indicators—such as physiological signals 
that reflect arousal before behavioral degradation \cite{Boucsein2012, Kosch2023}, and 
trigger interventions precisely when each specific respondent begins to struggle 
with each specific item.

We present an adaptive survey assistance system that addresses this gap by combining real-time ubiquitous sensing with personalized, dynamically updating prediction models. Our system uses electrodermal activity (EDA) and mouse movement data to continuously assess respondent cognitive state, predicting moment-to-moment when respondents need help with knowledge-based survey items. Critically, we employ personalized classifiers with threshold adaptation that align with each respondent's baseline patterns throughout the session. This enables the system to distinguish an individual's typical processing behavior from genuine difficulty, triggering LLM-based clarifications and explanations only when cognitive load indicators suggest the respondent is struggling. Unlike static systems that apply predetermined support schedules or reactive systems that wait for explicit help requests, our approach provides proactive interventions timed to each individual's actual cognitive state as it evolves across sequential items.

A within-subjects study (N=32) compared our aligned-adaptive (adjusted toward cognitive states) assistance timing against mis\-aligned-adaptive (adjusted backward cognitive states) and random-adaptive control conditions across sequential survey items containing factual knowledge questions. Results demonstrate that physiological and behavioral signals can effectively predict when respondents need help. The adaptive assistance improves response accuracy from 41\% to 62\% and reduces missed assistance opportunities compared to the control conditions. Participants rated the aligned-adaptive system significantly higher on efficiency, dependability, and benevolence scales, while the aligned-adaptive condition achieved the highest acceptance rate, indicating positive user acceptance of real-time intervention. 

Our findings highlight the critical role of timing in proactive assistance: interventions that arrive when respondents actually struggle yield greater benefits than those delivered too early or too late, or misaligned with individuals' real-time need. By preventing small difficulties from compounding across items, aligned-adaptive timing helps preserve both response quality and user experience. This work provides a foundation for adaptive survey systems that maintain data quality through physiological and behavioral sensing, with applications that span educational assessments, healthcare questionnaires, and research surveys where the complexity of the item creates a cumulative burden on the respondent.

\section{Related Work}
We discuss related work on (1) support elements in surveys, (2) proactive assistance with LLMs, and (3) physiological and behavioral sensing for cognitive states and its adaptive applications.

\subsection{Interactive or Intelligent Support in Survey Methodology} 
Survey researchers have long studied how task difficulty and motivation shape response quality. High task difficulty can lead to data issues, such as satisficing \cite{Roberts2019, Blazek2023}, speeding \cite{Cannell1981} and item nonresponse \cite{Haunberger2011}. Difficult questions also increase the risk of drop-out \cite{Liu2017, Hoerger2010}. Longer and more complex questions demand greater cognitive effort, which has been linked to higher breakoff rates even when other factors are controlled \cite{Peytchev2009}.

To address challenges in data quality and respondent experience, researchers have explored the use of interactive and pre-defined support within web surveys. Early studies focused primarily on reducing comprehension problems through system-initiated clarifications. These interventions, which included the use of hyperlinks to definitions, mouse-over explanations, and proactive messages triggered by long response times or pauses \cite{conrad2003interactive, schober2000clarifying, conrad2006use, Conrad2007}, were found that respondents answered more accurately when compared to surveys without such support. However, these methods primarily addressed pre-defined clarity rather than behavioral issues.

More recent work has expanded beyond clarification to  examine interactive feedback mechanisms designed to reduce satisficing, careless responding, and improve engagement. Studies have shown that providing respondents with feedback on speeding, reminders to read carefully, or commitment prompts encouraging conscientious answering can enhance response quality and improve accuracy by encouraging more thoughtful responses \cite{Kunz2019, https://doi.org/10.18148/srm/2017.v11i1.6304}. These approaches acknowledge that misunderstandings are common and cannot be eliminated through pretesting alone \cite{Conrad2000}. For example interactive probes can also elicit richer answers, especially from participants who are highly engaged with the survey topic \cite{Holland2008}. 

Despite these successes, a limitation persists across both early and contemporary approaches: they predominantly rely on uniform, ``one-size-fits all'' assumptions about respondent behavior, and do not account for inter-individual or moment-by-moment variation.  Most studies deploy static intervention triggers, such as a fixed response-time threshold to identify speeding, which fail to accommodate significant individual differences in reading speed, cognitive load, engagement, or interaction patterns.  Help can be beneficial when offered at the right moment, but distracting or even harmful when mistimed \cite{https://doi.org/10.18148/srm/2017.v11i1.6304}. This gap highlights the need for a more dynamic approach that tailors assistance to individual needs rather than relying on universal, predefined rules.  

Recent work has introduced LLMs as a flexible tool for intelligent survey support. LLMs can generate alternative question phrasings or conversational prompts to maintain engagement \cite{Yun2024, Mburu2025}. They can build glossaries or synonyms to lower cognitive barriers in domain-specific contexts \cite{VidalSabans2025}. They can also integrate contextual user data from online platforms to extend survey functionality \cite{Velykoivanenko2024}. Beyond LLMs, interactive question answering systems allow surveys to engage respondents in dialog, clarifying ambiguities, and improving accuracy \cite{Biancofiore2024}. These developments show the potential of intelligent, context-aware systems to enhance survey quality, but they mainly rely on textual interaction and static behavioral cues, leaving moment-to-moment cognitive variation unaddressed. 

Our work complements and extends this line of research by incorporating ubiquitous and non-intrusive sensing to capture real-time changes in cognitive load, enabling personalized, proactive support that aims to be delivered precisely when respondents need it. This focus on adaptive timing and individualized state estimation fills a critical gap in existing approaches and illustrates how multimodal sensing can further advance intelligent survey interaction.
 
\subsection{Proactive Assistance with LLMs}

Proactive assistance systems aim to provide support before users explicitly ask 
for it. Early research introduced the idea of Just-In-Time Information Retrieval 
agents (JITIRs), which proactively present useful information based on local 
context in a non-intrusive way \cite{Rhodes2000}. These systems reduce the 
cognitive effort of searching and encourage users to access information they 
might otherwise miss. Other work explored proactive agents that listen to ongoing 
conversations, detect key entities, and retrieve related information, thereby 
reducing the need for explicit search activity \cite{Andolina2018}. These studies 
established proactive support as a way to reduce mental effort and improve task performance.

Designing proactive AI assistants that deliver positive experiences remains 
challenging, since effective collaboration principles differ across tasks and 
contexts. Recent work has explored proactive assistance with LLMs across a range 
of domains, highlighting both the opportunities and challenges of this approach. 

In care and wellbeing contexts, Liu et al. \cite{Liu2024} 
developed ComPeer, a conversational agent for proactive peer support that detects 
significant emotional events in conversations and strategically times interventions. Across 18 users over two weeks, they found that timing accounted for 40\% of variance in intervention acceptance—identical suggestions were accepted three 
times more often when delivered at moments users perceived as appropriate (e.g., 
after expressing frustration) versus inappropriate moments (e.g., mid-task). 
Their finding that ``timing matters more than content quality'' directly motivated 
our focus on temporal alignment. Building on these works, our study explores 
proactive LLM assistance in the context of cognitive overload in survey-like interaction.

In programming contexts, Chen et al. \cite{Chen2025Need} explored whether 
LLMs could proactively anticipate developers' needs and offer code suggestions 
before being asked. Across 24 developers, they found preemptive assistance 
improved efficiency by 17\% when timed to natural workflow pauses (e.g., after 
completing a function), but risked learned helplessness when mistimed—developers 
began waiting for AI help rather than attempting problems independently when 
suggestions appeared too frequently. This highlights that adaptive timing must 
balance support with user agency.

\citet{Pu2025} explicitly evaluated ``assistance versus disruption'' 
tradeoffs in proactive programming support through 24 developer interviews. They 
identified ``temporal appropriateness'' as the primary factor distinguishing 
helpful from disruptive interventions—the same code suggestion was perceived as 
supportive when timed to natural pause points but disruptive when interrupting 
active coding, even if content quality was identical. They found that developers 
strongly preferred systems that could detect their cognitive state rather than 
intervening on fixed schedules.

In healthcare, Imrie et al. \cite{imrie2023redefining} developed proactive LLM-based 
explanations for medical terminology in patient-facing health forms. Their system 
monitored when patients hovered over unfamiliar terms and automatically generated 
plain-language explanations. Across 89 patients completing health history forms, 
proactive explanations reduced completion time by 18\% and improved accuracy of 
self-reported symptoms compared to on-demand help buttons. However, their trigger 
remained behavioral—dwell time exceeding 2 seconds—requiring users to first 
encounter and visibly pause on problematic content before receiving help. This 
highlights the limitation of behavioral triggers: they detect struggle only after 
it has begun manifesting in observable behavior.

These studies converge on a key insight: \textbf{timing matters as much as 
content}. However, all rely on behavioral triggers, pause detection \cite{imrie2023redefining}, 
workflow stage transitions \cite{Chen2025Need, Pu2025}, or conversational cues 
\cite{Liu2024}, which are inherently retrospective. Behavioral indicators appear 
only after cognitive processes have begun to affect observable actions. Our work 
extends this by exploring whether \emph{physiological} and \emph{behavioral} triggers can enable even 
earlier, more prospective intervention. By detecting cognitive load through EDA before it manifests in behavior, we test whether intelligent support can be timed to the moment the struggle begins rather than after it becomes behaviorally apparent.
 
\subsection{Physiological and Behavioral Sensing for Cognitive State Detection and Adaptive Applications} 

Physiological and behavioral sensing has become a central approach for detecting cognitive states such as workload, stress, and fatigue, enabling adaptive systems to provide timely support. A broad range of physiological signals have been explored for this purpose. For example, eye tracking and electrocardiograms (ECG) have been shown to capture fluctuations in cognitive load and stress in virtual reality (VR) environments, allowing for real-time adaptation \cite{Nasri2025, Gao2024}. Electroencephalography (EEG) remains a widely used modality, with recent advances demonstrating that even consumer-grade, headphone-style EEG devices can provide reliable signal quality and classification of cognitive load across diverse tasks when configured by trained experimenters \cite{Knierim2025, Kohlmorgen2007}. Similarly, multi-sensor approaches including ECG, EEG, EDA, and facial expressions have been applied to assess cognitive status, such as attention, fatigue, and workload during human–robot collaboration and human-human interview scenarios \cite{Jaiswal2024,  Lim2021, 10756706}. Planke et al. \cite{Planke2021} further showed that online multimodal fusion of EEG, eye activity, and control inputs yields accurate and reliable inference of mental workload in driving scenarios. Among these signals, EDA is particularly prominent due to its sensitivity to sympathetic nervous system activation and its established use in workload detection and user experience evaluation \cite{Boucsein2012, Kosch2023, Georges2016, Schaule2018}. However, not all physiological measures are equally suitable for short interaction windows; for instance, cardiovascular metrics typically require sustained engagement (over one minute) to reliably reflect cognitive load \cite{shaffer2017overview, electrophysiology1996heart}. 

Complementing these physiological measures, behavioral sensing methods offer lightweight and scalable alternatives. Eye-tracking features such as fixation dynamics, combined with heart rate data, have been used to differentiate low and high cognitive load states in participant-specific models in gaming \cite{Appel2019, Appel2023}. More recently, mouse tracking has emerged as a promising proxy for cognitive processes in digital environments. Unlike simple response time measures, mouse trajectories reflect the continuous evolution of decision-making, capturing hesitation, uncertainty, and conflict in real time \cite{cisek2010neural, freeman2018doing}. This approach is especially suitable for online tasks, as it is virtually cost-free, scalable, and more robust to external distractions compared to traditional latency-based metrics \cite{horwitz2020learning, horwitz2017using, Dias2019PredictingRU}. Features such as repeated directional changes and prolonged hovering of the cursor have been associated with an increased response burden in survey tasks, providing fine-grained indicators of cognitive strain \cite{horwitz2020learning, leipold2024detecting, fernandez2023predicting}. These findings underscore the potential of integrating physiological and behavioral sensing for real-time, fine-grained modeling of cognitive states across diverse application domains.

While these advances demonstrate the potential of multimodal sensing for adaptive support, they come almost exclusively from domains such as VR training, gaming, driving, robotics, and learning technologies, not from survey methodology. Survey research has historically relied on static interfaces and fixed rules (e.g., response-time thresholds, scripted clarification prompts), with very limited use of moment-to-moment behavioral or physiological monitoring. Thus, there is a clear gap: surveys rarely incorporate adaptive sensing to infer cognitive state, even though such techniques are well established in other domains.

Our work bridges this gap by drawing on adaptive sensing practices from HCI while grounding the application in survey methodology. We combine lightweight physiological and behavioral signals to build an LLM-based agent capable of dynamically tailoring support, offering help precisely when cognitive load is elevated, addressing challenges that traditional survey designs cannot capture.





\section{Design of the Adaptive System for Surveys with a Proactive AI Agent}
We designed an LLM-based AI agent that proactively intervenes to mitigate cognitive overload while users complete web-based multiple-choice questions. 

\subsection{Feature Selection and Baseline Classifiers}
We implemented a system that detects overload and proactively assists users.  Cognitive overload detection is enabled by continuously processing physiological and behavioral data to classify users’ cognitive states and dynamically trigger intervention; this classifier is further personalized and adapted throughout interaction. Further, LLM-based help for cognitive overload mitigation delivers proactive, task-related assistance when overload is detected, combining interactive clarifications and explanations that reduce cognitive strain and support task completion.

We used a dataset \cite{liu2026physiologicalbehavioralmodelingstress} from a web-based multiple-choice task with manipulated question difficulty, focusing on general knowledge questions. Multimodal data were collected, including mouse dynamics, eye tracking, ECG, and EDA.

To identify the most informative features to indicate cognitive load, we applied \texttt{SelectKBest} with \texttt{f_regression} as the scoring function. We chose \texttt{f_regression} because it evaluates linear correlations between features and the target variable, aligning with our use of linear regression models for baseline prediction. The results (cf. \autoref{sec:featureimp}) indicated that the top predictive features included three mouse movement–based measures and two EDA measures: the number of vertical direction changes in cursor movement (\textit{ypos_flips}), the duration of time the cursor hovered over (\textit{hover_time}), the total number of hovering events during a question (\textit{hovers}), the number of peaks (\textit{peaks_num}) and the average tonic EDA level across the task (\textit{tonic_avg}). We decided to focus on mouse movement and tonic EDA measure because they can be computed robustly in real time. We excluded features unsuitable for real-time interaction. Although peak-based EDA features (\textit{peaks_num}) ranked highly offline, they require windowed peak detection and smoothing, introducing latency. Because f-regression–based feature selection does not alter the ranking of suitable features when unsuitable ones are excluded, the final feature set balances predictive power with low-latency, stable signals for live adaptation. Based on these results, we trained two baseline regression models to estimate cognitive load: one using tonic EDA and one using mouse-movement features. To reduce individual and task-related baseline effects, the EDA model used changes from task-onset baseline rather than absolute values and incorporated task difficulty (binary-coded). The mouse-based model combined cursor direction changes, hover duration, number of hover events, and task difficulty in a linear regression. We included pre-validated task difficulty \cite{liepmann2010bowit} as a feature because easy and difficult questions were intentionally balanced and randomly interleaved. This provided relevant context for distinguishing genuine cognitive overload from responses to question difficulty alone, preventing threshold drift and over-triggering of assistance on subsequent easy questions. Importantly, task difficulty serves only as a minor contextual predictor; it does not dominate 
$y_{\text{EDA}}$ or $y_{\text{Mouse}}$ on its own. When EDA or mouse-movement patterns remain stable, the model output stays low and does not trigger adaptation.
\[
y_{\text{EDA}} = \alpha \cdot \text{tonic\_difference} + \beta \cdot \text{task\_difficulty} + \epsilon
\]
\[
y_{\text{Mouse}} = \alpha \cdot \text{ypos\_flips} + \beta \cdot \text{hover\_time} + \gamma \cdot \text{hovers} + \delta \cdot \text{task\_difficulty} + \epsilon
\]

We used two separate unimodal models to increase robustness, allowing one modality to provide stable estimates if the other is degraded by noise, connectivity issues, or missing data.

\begin{figure*}[htbp]
    \centering
    \includegraphics[width=1.0\textwidth]{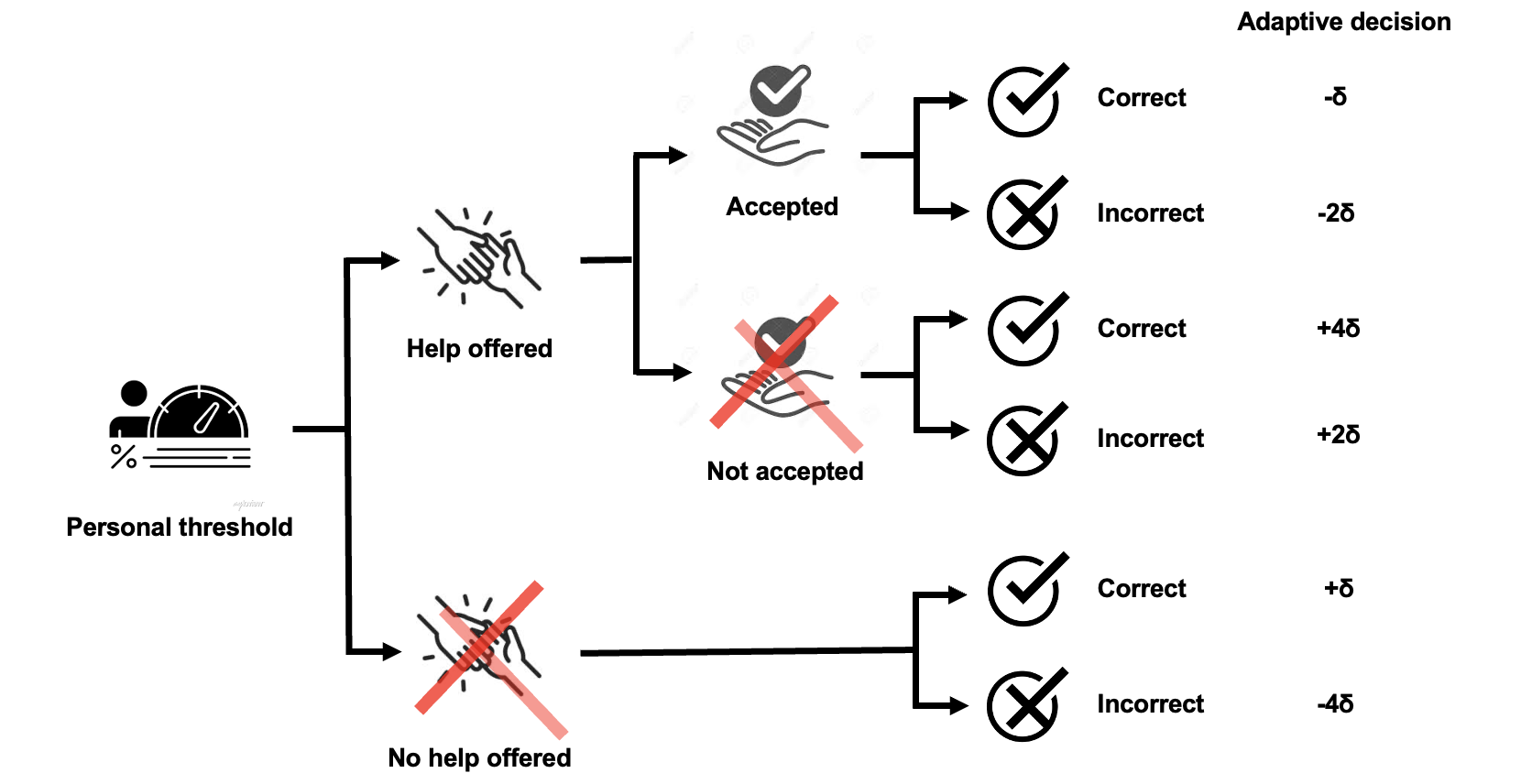} 
    \caption{Rule-based threshold adaptation logic per trial: positive changes increase the threshold (fewer future interventions), negative changes decrease it (more future interventions).}
    \Description{Decision tree indicating the threshold adaption based on whether help was offered and accepted, and whether the respondent's answer was correct.
    Root: personal threshold.
    Traversal 1: Help offered --> help accepted --> correct answer --> threshold adaption by -sigma
    Traversal 2: Help offered --> help accepted --> incorrect answer --> threshold adaption by -2*sigma
    Traversal 3: Help offered --> help not accepted --> correct answer --> threshold adaption by +4*sigma
    Traversal 4: Help offered --> help not accepted --> incorrect answer --> threshold adaption by +2*sigma
    Traversal 5: No help offered --> correct answer --> threshold adaption by +sigma
    Traversal 5: No help offered --> incorrect answer --> threshold adaption by -4*sigma.}
    \label{fig:threshold-rules}
\end{figure*}

\subsection{Personalization and Adaptive Process}
To enable personalization and adaptive support, we extended the baseline classifiers through both gradient-based fine-tuning and rule-based adaptive updates. The model of each user was personalized using gradient descent with L2 regularization, ensuring stable convergence and preventing overfitting. At the beginning of the interaction, users completed a calibration phase consisting of self-reported cognitive load questions. These calibration responses were used to fine-tune the model intercept ($\epsilon$) and feature weights through a one-shot gradient-style update, thereby aligning the system’s predictions with individual’s subjective perception of cognitive effort.
The final output was computed as the maximum of the EDA-based and mouse movement–based classifiers:
\[
y_{\text{Final}} = \max(y_{\text{Mouse}}, y_{\text{EDA}})
\]
This design prioritizes sensitivity, ensuring that potential overload detected by either modality is not overlooked. To trigger interventions, we defined a decision threshold 
$\theta$, which kept rule-based updating after each experimental question, i.e., $\theta_{current}=\theta_{previous}\pm k\delta$, $k$ was determined by users' interaction. If $y_{\text{Final}}$ > $\theta$, the system classified the user as experiencing cognitive overload and proactively offered help; if $y_{\text{Final}} \leq \theta$, no intervention was triggered. Both $y_{\text{EDA}}$ and $y_{\text{Mouse}}$ were implicitly aligned through the initialization and personalization procedure.

Further, the system adapted dynamically during task through rule-based threshold adjustments. These rules were grounded in two principles: (1) missed opportunities to intervene should lower the threshold to catch future overload earlier, and (2) unnecessary or unwanted interventions should raise the threshold to avoid over-assistance. These updates were guided by the outcome of each task attempt and the user’s response to offered help:
\begin{itemize}
\item \textbf{No help offered, answer correct}: the threshold was slightly increased (+$\delta$) as the user successfully managed the task independently, suggesting earlier intervention was unnecessary.
\item \textbf{No help offered, answer incorrect}: the threshold was substantially decreased (-4$\delta$) because the system missed an opportunity to intervene; it should step in earlier in future.
\item \textbf{Help offered, accepted, and answer correct}: the threshold was slightly decreased (-$\delta$), reflecting that the help was useful and should be offered somewhat earlier.
\item \textbf{Help offered, accepted, and answer incorrect}: the threshold was moderately decreased (-2$\delta$), as the user sought help, but the support was insufficient; earlier and possibly stronger intervention may be needed.
\item \textbf{Help offered, not accepted, and answer correct}: the threshold was substantially increased (+4$\delta$), as the user demonstrated that they did not need assistance, indicating the system should hold back more.
\item \textbf{Help offered, not accepted, and answer incorrect}: the threshold was moderately increased (+2$\delta$), since the user rejected the help but still performed poorly, suggesting reluctance to accept intervention and that the system should offer it less frequently.
\end{itemize}

Through this gradient-based personalization combined with rule-based adaptation, the system continuously refined its intervention strategy to better align with each user’s behavior. We summarize the decision rules in \autoref{fig:threshold-rules}.

\subsection{Development and Implementation}
\label{sec:implementation}
The system was implemented in Python and JavaScript with a Flask-based backend that coordinated real-time data collection, classification, and interaction. Mouse movement data was captured from the browser and EDA data collected via Bluetooth. Both streams were processed in real time, and the classifier dynamically evaluated the user’s cognitive state at each trial. 

When the classifier detected cognitive overload and crossed the intervention threshold, the system triggered a pop-up prompt asking whether the user would like assistance. If the user declined or ignored, the task continued uninterrupted. If the user accepted assistance, the interface allowed them to freely select any portion of the question text to query, ranging from a single word to a phrase to the entire question. We designed this interaction to maximize user control and flexibility. Instead of imposing predefined highlights or system-selected segments, users could indicate precisely where they experienced uncertainty. It reduces the risk of unnecessary or irrelevant explanations, ensuring that system-provided help is both context-sensitive and user-driven. The selected text was then sent to the backend for processing (cf. \autoref{fig:study-screenshots} (d)-(f)).

For providing explanations, we embedded a locally hosted large language model (LLaMA-2-7B \cite{touvron2023llama2openfoundation}). Upon receiving the selected text, the model generated a natural language explanation tailored to the user’s query (using a fixed prompt cf. \autoref{sec:prompt}) and its responses were limited to a maximum of 160 tokens.

\section{User Study}

Our study investigates how adaptive LLM assistance timing affects survey response quality and user experience in sequential knowledge tasks. We engaged participants in multiple-choice questions while continuously monitoring physiological and behavioral signals to trigger real-time interventions. 

We used a three-condition within-participants experimental design with the independent variable \textsc{Adaptation Strategy} (\textit{Aligned-Adaptive}, \textit{Misaligned--Adaptive}, and \textit{Random-Adaptive}). \textsc{Adaptation Strategy} describes how the system adjusts intervention thresholds based on user performance and help acceptance behavior. The \textit{Aligned-Adaptive} condition aligns threshold updates with user needs, \textit{Misaligned-Adaptive} deliberately misaligns them, and \textit{Random-Adaptive} applies arbitrary adjustments. All conditions provide identical LLM-based assistance content, differing only in timing mechanisms. The study aimed to test central hypotheses:
\begin{itemize}
   \item \textbf{H1 [Task Performance]}: Response accuracy will be significantly higher in the \textit{Aligned-Adaptive} system than in the \textit{Misaligned-Adaptive} or \textit{Random-Adaptive} systems.  
   \item \textbf{H2 [User Experience]}: The \textit{Aligned-Adaptive} system will lead to superior perceived user experience, reflected in higher ratings of efficiency, dependability, and benevolence, compared to the \textit{Misaligned-Adaptive} or \textit{Random-Adaptive}.  
   \item \textbf{H3 [Workload]}: Participants will report lower subjective workload when interacting with the \textit{Aligned-Adaptive} system, compared to the \textit{Misaligned-Adaptive} or \textit{Random-Adaptive}.  
\end{itemize}

A total of 32 participants completed the study using a standard personal computer. The study was approved by the ethics committee of LMU Munich (EK-MIS-2025-0414-FT-d01).

\subsection{Task}

The experimental task consisted of answering multiple-choice questions from validated German-language cognitive test batteries \cite{liepmann2010bowit}, subsequently translated into English. Each question had five options, only one of which was correct. To ensure a structured manipulation of task difficulty, questions with a reported average accuracy rate above 60\% in the tests' population norm were categorized as easy, while those with a reported average accuracy rate below 60\% were categorized as difficult.

The questions spanned a diverse set of general knowledge domains, including Fine Arts, Biology/Chemistry, Health, Geography, History, Politics, Mathematics, Religion, Literature, and Technology. We prepared four blocks of 20 questions each. Within each block, the distribution of easy and difficult questions was counterbalanced to ensure comparable overall difficulty across blocks. Topical coverage was balanced to prevent any single domain from disproportionately influencing performance.

\subsection{Study Design}
The experiment employed a within-subjects design in which each participant experienced three distinct conditions, along with a calibration block presented at the very beginning. The calibration block consisted of the same number of questions as the main experimental blocks and was used to fine-tune the personalized classifier for each user based on self-reported cognitive load in a 7-point scale. Each of the three main conditions, \textit{Aligned-Adaptive}, \textit{Misaligned-Adaptive}, and \textit{Random-Adaptive}, manipulated the intervention threshold mechanism differently to examine its effect on user interaction, cognitive load, and task performance. To control for order effects, the sequence of the three experimental conditions and the assignment of question blocks to the conditions was randomized across participants. Furthermore, the order of questions within each block was randomized independently. The starting classifiers of all three conditions were the same, with the same feature weights and intercepts. 

Based on a two-participant pilot study, we set the initial threshold to $\theta = 12$, requiring the estimated overload level to exceed this value before assistance was triggered. The step size was set to $\delta = 1$ to allow gradual threshold adaptation across conditions.

\subsubsection{Aligned-Adaptive} In this condition, the intervention threshold followed the rule-based adaptation strategy described above \autoref{fig:threshold-rules}. The threshold was dynamically adjusted after each question based on the user’s performance and response to offered help, allowing the system to provide contextually timed interventions.
\subsubsection{Misaligned-Adaptive} The threshold adjustments were inverted relative to the standard rule. For example, increases in the threshold in the standard rule became decreases, and vice versa. This condition tested the sensitivity of the system’s adaptation logic and its impact on user experience.
\subsubsection{Random-Adaptive} The threshold was adjusted by a randomly selected value within the same absolute range as the adaptive rule. For example, if the maximum change in the adaptive condition was ±x, the condition applied a random adjustment between –x and +x after each question trial. This condition served as a control to assess the effects of structured adaptation compared to unstructured, stochastic threshold changes.

\subsection{Apparatus}
The experimental setup integrated a combination of web-based task presentation, mouse movement tracking, and physiological data collection, through a Chrome browser window running our custom-built Flask web application (cf. \autoref{sec:implementation}). 

\subsection{EDA Recording and Processing}
EDA was continuously recorded using the BITalino (r)evolution kit (PLUX Wireless Biosignals, Portugal) at a sampling rate of 100 Hz, with \texttt{revolution-python-api} provided by BITalino. To ensure high signal quality, two electrodes were attached to the index and middle fingers of the participant’s non-dominant hand, minimizing interference with the mouse-hand used for task interaction. Prior to electrode placement, the skin was cleaned and treated with a potassium chloride (KCl) solution to reduce impedance and enhance conductivity. A custom Python monitoring module was developed to manage data acquisition, segmentation, and storage in real time. It also time-aligned the physiological signals with task events.

The monitoring module maintained two levels of data handling: (1) session-level logging, where all EDA samples were continuously appended to a cumulative session file, and (2) question-level segmentation, where EDA signals were stored separately for each trial to enable fine-grained analysis. To reduce data loss in case of technical errors, periodic backups were automatically triggered at one-minute intervals. Each file was indexed by user ID, question index, and timestamp for traceability.

For each trial question, the system extracted the initial tonic level of the EDA signal at each question onset. Subsequent fluctuations were computed relative to this baseline.

During acquisition, a dedicated acquisition thread streamed raw EDA data, time-stamped each reading, and associated it with both the local question index and global session index. This ensured that EDA features could be aligned with behavioral markers such as mouse movements. At the end of each question and upon session termination, data were automatically finalized and stored.

\subsection{Mouse Movement Recording and Processing}

Mouse movement data were continuously recorded within the web-based task interface using a custom JavaScript tracking module. 
The tracker monitored and logged several features on a per-trial basis. One key measure was the direction flip count, defined as the number of times vertical cursor movement changed direction after exceeding a displacement threshold of 100 pixels, which served as a proxy for disorganized or uncertain exploration. Another was the hover count, measuring how often the cursor remained stationary for longer than 500 ms, reflecting moments of hesitation or focused inspection. In addition, the system recorded the total hover time, capturing the cumulative duration of all hover events within a trial as an indicator of prolonged deliberation.

To ensure temporal alignment with task events, each mouse movement was timestamped and linked to both the local question index and the global session index. For each trial, a summary log was created containing total flips, hover events, and hover durations, which was then transmitted asynchronously to the Flask backend via REST API calls. Intermediate data (e.g., ongoing hovers) were finalized at the end of each trial, ensuring that no interaction events were lost during transitions.

\subsection{Measures}
We evaluated both user-related outcomes and system-related outcomes, combining objective task performance, self-reported perceptions, and system-level metrics. This approach provided a comprehensive understanding of how participants interacted with the system and how well the system functioned.

User-related outcomes focused on \textbf{Task Performance}, \textbf{Task Effort}, and \textbf{User Experience (Benevolence, Efficiency, and Dependability)}.
System-related outcomes focused on classifier performance, quantified using confusion matrices to analyze detection accuracy, false positives, and false negatives.

\subsubsection{Task Performance and Effort}
Objective task performance was measured by the number and percentage of correctly answered multiple-choice questions. To complement behavioral outcomes, we assessed participants’ subjective perceptions of workload and performance using the unweighted NASA-TLX \cite{Hart1988}. 

\subsubsection{Benevolence}
We further examined participants’ perceptions of the system’s benevolence, defined as the belief that the system acts in the user’s best interest and genuinely supports their goals \cite{Gulati2019}. We used the ``benevolence'' subscale of the human computer trust scale \cite{Gulati2019, Dubiel2024} in 5-point Likert items: 
\begin{itemize}
\item \textbf{Interest}: I believe that the system will act in my best interest.
\item \textbf{Help}: I believe that the system will do its best to help me if I need help.
\item \textbf{Preferences}: I believe that the system is interested in understanding my needs and preferences.
\end{itemize}

\subsubsection{Efficiency and Dependability}
User experience was measured using the modular UEQ+, an extension of the original User Experience Questionnaire (UEQ) \cite{Laugwitz2008}. As per the authors' recommendations for partial deployment, we focused on the two scales most relevant to the system’s interaction design:
\begin{itemize}
\item \textbf{Efficiency}: tasks can be completed without unnecessary effort. This scale contrasts items, i.e., slow/fast, inefficient/efficient, impractical/practical, and cluttered/organized.
\item \textbf{Dependability}: the user remains in control of the interaction. This scale contrasts items, i.e., unpredictable/predictable, obstructive/supportive, not secure/secure, and does not meet expectations/meets expectations.
\end{itemize}

\subsubsection{System Performance}
To evaluate the accuracy of the system’s cognitive overload detection, we collected ground-truth labels directly from participants during each question trial. After attempting a question, participants were asked whether they felt they needed help, responding with one of two options: Yes or No. These self-reported judgments served as an independent measure of perceived task difficulty and were not tied to the functionality of the AI assistance trigger. 

By comparing the predicted overload states of the model with these independent self-reports, we constructed confusion matrices to assess detection accuracy, false positives, and false negatives in post-hoc validation. We further measured the number of triggered assistance and the number of accepted assistance.

\begin{figure}[htbp]
    \centering
    \includegraphics[width=0.49\textwidth]{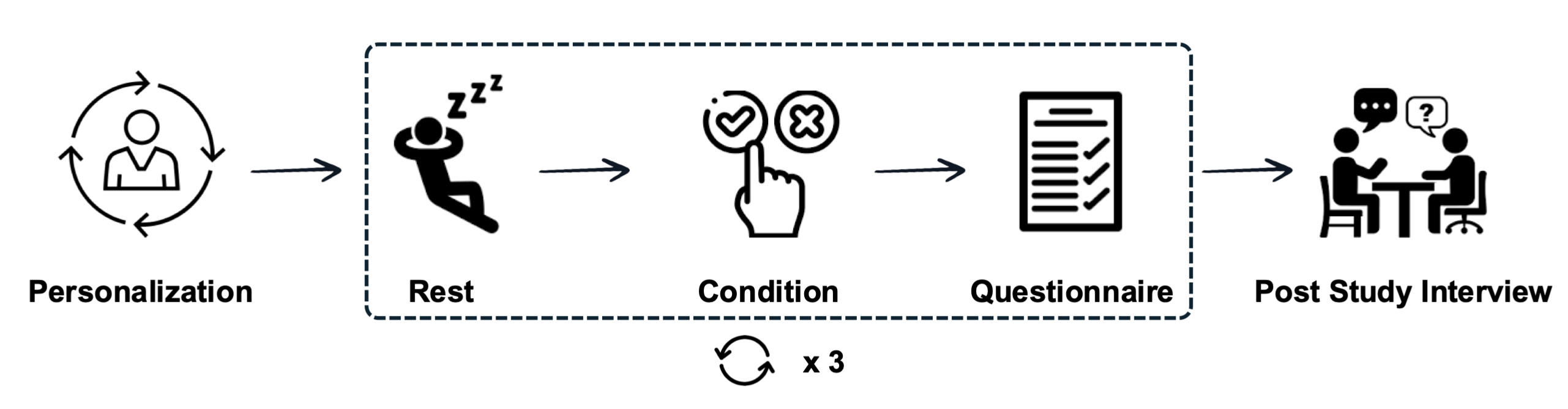} 
    \caption{The task was structured in blocks. First, participants completed a personalization block to calibrate physiological and behavioral signals. This was followed by three condition blocks, \textit{Aligned-Adaptive}, \textit{Misaligned-Adaptive}, and \textit{Random}, presented in randomized order (×3). After each condition block, participants provided evaluation ratings, and the session concluded with a post-study interview.}
    \Description{Flow diagram of study design. Step 1: Personalization. Step 2: Block of Step 2.1: Rest, Step 2.2: Condition, Step 2.3: Questionnaire. Steps 2.1 to 2.3 are repeated 3 times. Step 3: Post-study interview.}
    \label{fig:overview}
\end{figure}

\subsection{Procedure}
Upon arrival, participants were welcomed by the experimenter and provided with a comprehensive introduction to the study. This included an overview of the research goals, the role of the intelligent system, and the types of data being recorded (mouse movement, EDA, task responses, and self-reports). Participants received detailed instructions on the procedure and system interaction. They were informed that the AI might occasionally offer help, but were not told how assistance was triggered or about the experimental conditions. Participants could accept, decline, or ignore offers. Written informed consent was obtained, followed by a brief demographic questionnaire on age and gender.

Participants completed general knowledge multiple-choice questions using a mouse while the system continuously monitored cognitive state and triggered assistance prompts when thresholds were exceeded. When help was accepted, participants could select any portion of the question text and submit it for clarification, and were instructed on this process at the start of the study.

The session comprised four blocks (\autoref{fig:study-screenshots},  \autoref{fig:overview}): a calibration block to personalize the classifier using self-reported cognitive load, followed by three experimental blocks, \textit{Aligned-Adaptive}, \textit{Misaligned-Adaptive}, and \textit{Random-Adaptive}. Experimental block and question orders were randomized across participants. Each block included 20 questions, with mandatory breaks of at-least one-minute between blocks to reduce fatigue. After each experimental block, participants completed a questionnaire on workload, user experience, and system perceptions.

At the conclusion of the session, participants engaged in a semi-structured interview (cf. \autoref{sec:interview}). During the interview, they were asked to reflect on their overall experience with the system, describe their perceptions of the adaptive interventions, and suggest potential improvements for future iterations. The entire experiment took around one hour.

\subsection{Participants}
An a priori power analysis was conducted using \textit{G*Power} (version 3.1) \cite{Faul2009}, to estimate the required sample size for a study analyzed with linear mixed-effects models (LME) \cite{Faul2009}. The analysis assumed a medium effect size ($f^2 = 0.25$) based on HCI guidelines \cite{Yatani2016}, an alpha level of $0.05$, and a desired power of $0.80$. The results indicated that a minimum total sample size of $27$ participants is required to achieve a desired statistical power (actual power = $0.81$). A total of 32 valid participants took part in the study, recruited from a university through message groups. The sample included $14$ males and $18$ females. Participants' ages ranged from $19$ to $43$ years (M = $26$, SD = $5.77$). The participants received a flat payment of 12 euros.

\section{Metadata and Validation}
To contextualize the empirical findings, we report metadata related to study conditions and system performance. 
\subsection{Condition Duration}
Participants spent the following amounts of completion time in each condition.  
For the \textit{Aligned-Adaptive} condition, the mean duration was 17.71 minutes (SD = 13.08); the \textit{Random–Adaptive} condition, 14.62 minutes (SD = 13.52); the \textit{Misaligned–Adaptive} condition, 18.36 minutes (SD = 16.55). A one-way ANOVA was conducted to examine differences in task durations across the three conditions. The analysis could not detect a statistically significant effect of condition on completion time ($p = 0.56$).

\subsection{System Performance}
\subsubsection{Performance to know when people need help}
\begin{figure*}[htbp]
    \centering
    \includegraphics[width=1.0\textwidth]{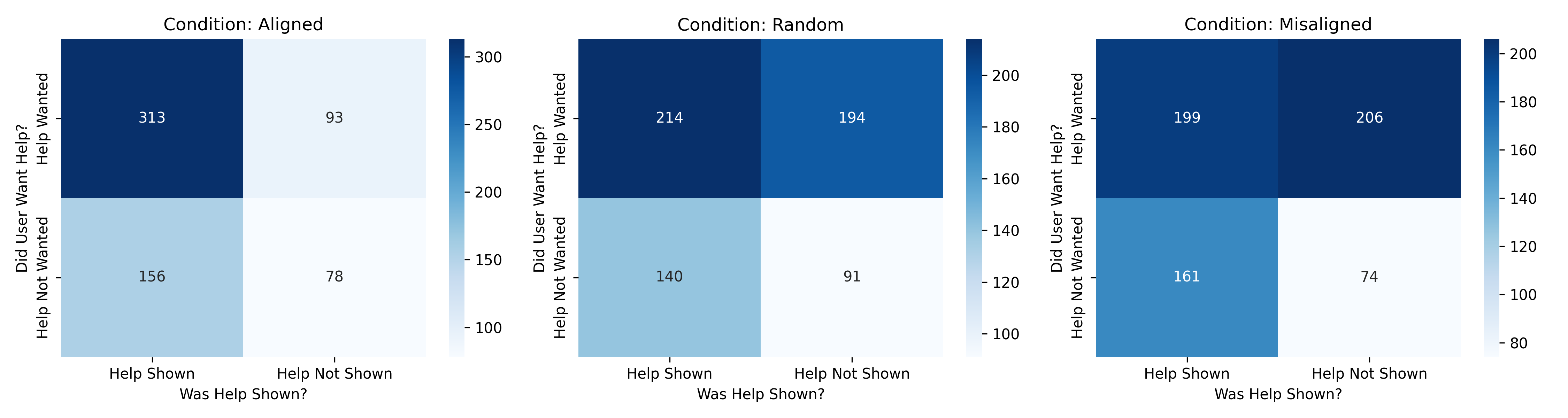} 
    \caption{Confusion matrices showing model performance in knowing whether users need assistance across the three experimental conditions (\textit{Aligned-Adaptive}, \textit{Misaligned-Adaptive}, and \textit{Random-Adaptive}). Each matrix compares the system's predicted cognitive state with the participants' self-reported need, illustrating accuracy, false positives, and false negatives per condition.}
    \label{fig:acc}
    \Description{Three confusion matrices showing help shown vs. not shown on the x-axis and help not wanted and help wanted on the y-axis.
    Matrix 1: Aligned condition. Help shown and not wanted: 156. Help shown and wanted: 313. Help not shown and not wanted: 78. Help not shown and wanted: 93.
    Matrix 2: Random condition. Help shown and not wanted: 140. Help shown and wanted: 214. Help not shown and not wanted: 91. Help not shown and wanted: 194.
    Matrix 3: Misaligned condition. Help shown and not wanted: 161. Help shown and wanted: 199. Help not shown and not wanted: 74. Help not shown and wanted: 206.}
\end{figure*}

As shown in \autoref{fig:acc}, in evaluating the adaptive regression model’s ability to detect participants’ needs for assistance, clear differences emerged across conditions. The \textit{Aligned-Adaptive} condition achieved the highest overall accuracy, averaging 61\%, indicating that the system could most reliably capture participants’ actual needs. In contrast, accuracy dropped to 48\% in the \textit{Random-Adaptive} condition and further to 43\% in the \textit{Misaligned-Adaptive} condition, reflecting substantially weaker performance in both sensitivity and specificity. These results suggest that only in the \textit{Aligned-Adaptive} condition achieved a sufficiently robust alignment with participants’ expressed help requests, whereas the \textit{Random-Adaptive} and \textit{Misaligned-Adaptive} conditions yielded less dependable outcomes.

To further examine errors, we compared false negative rates (FNR; proportion of missed help requests) across conditions at the participant level. This participant-level analysis provides a more fine-grained view than the aggregate confusion matrix, which collapses across all trials and participants. On average, the \textit{Aligned-Adaptive} condition exhibited a substantially lower FNR (M = 21\%, SD = 25\%) compared to both the \textit{Misaligned-Adaptive} condition (M = 44\%, SD = 50\%) and the \textit{Random-Adaptive} condition (M = 43\%, SD = 47\%). Paired Wilcoxon signed-rank tests with Holm-Bonferroni correction confirmed that these differences were statistically significant: \textit{Aligned-Adaptive} vs. \textit{Misaligned-Adaptive} (W = 16.0, p$_{corrected}$ = 0.0084) and \textit{Aligned-Adaptive} vs. \textit{Random} (W = 13.0, p$_{corrected}$ = 0.0052). These results indicate that the adaptive rule set was particularly effective at reducing false negatives, ensuring that user requests for help were less likely to go undetected compared to the \textit{Random-Adaptive} and the \textit{Misaligned-Adaptive} conditions.

\subsubsection{Helper Frequency and Distribution} 
\begin{figure*}[htbp]
    \centering
    \includegraphics[width=1.0\textwidth]{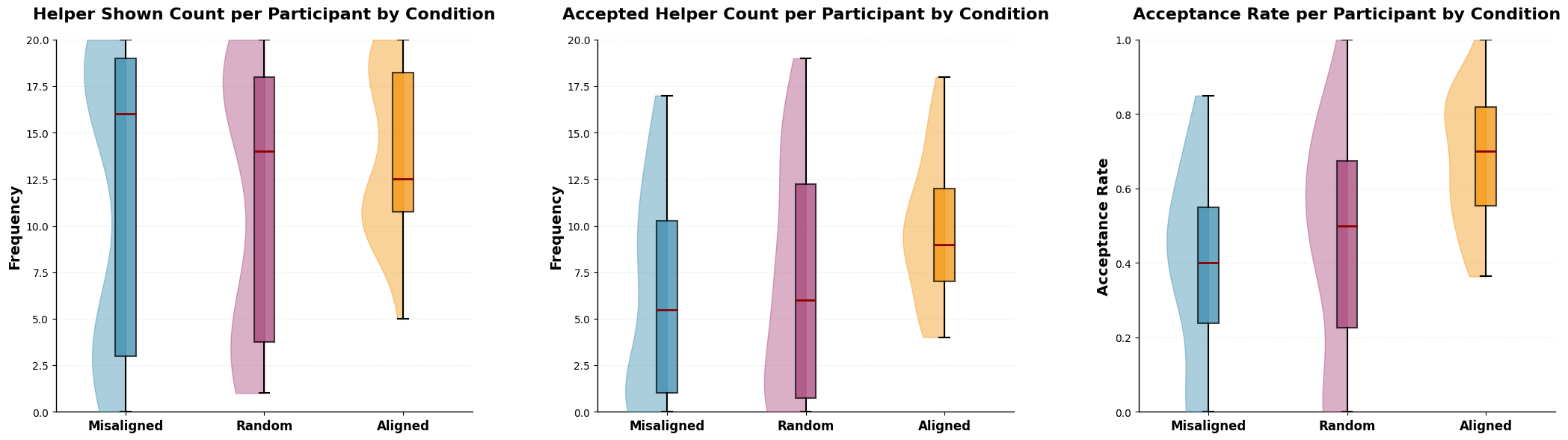} 
    \caption{Box plots showing rates of helpers (1) shown, (2) accepted, and (3) acceptance rate distributions across the three experimental conditions (\textit{Aligned-Adaptive}, \textit{Misaligned-Adaptive}, and \textit{Random-Adaptive}).}
    \Description{Three boxplots indicating how often the helper was shown and accepted as well as the resulting acceptance rate per condition.
    Plot 1: Helper shown count per participant. Misaligned: MD = 16 with a long tail towards lower counts. Random: MD = 14, also with a long tail towards lower counts. Aligned: MD = 12.5 with a narrower spread
    Plot 2: Accepted helper count per participant. Misaligned: MD = 5.5, Random: MD = 6, Aligned: MD = 9. Again, Aligned shows the smallest variance.
    Plot 3: Acceptance rate per participant. Misaligned: MD = 40\%, Random: MD = 50\%, Aligned: 70\%. Again, Aligned shows the smallest variance.}
    \label{fig:frq}
\end{figure*}

As shown in \autoref{fig:frq}, we analyzed helper presentation frequency, acceptance counts, and acceptance rates across conditions. The \textit{Aligned–Adaptive} condition showed the highest mean number of helpers (mean = 14.03, SD = 4.30, median = 12.50), compared to \textit{Misaligned–Adaptive} (mean = 11.50, SD = 7.76, median = 16.00) and \textit{Random–Adaptive} (mean = 11.19, SD = 7.17, median = 14.00). A similar pattern emerged for accepted helpers, with \textit{Aligned–Adaptive} producing more accepted helpers (mean = 9.75, SD = 3.76, median = 9.00) than \textit{Misaligned–Adaptive} (mean = 6.03, SD = 5.50, median = 5.50) or \textit{Random–Adaptive} (mean = 6.78, SD = 6.11, median = 6.00). Acceptance rates mirrored these trends, with \textit{Aligned–Adaptive} achieving the highest mean rate (mean = 0.676, SD = 0.171, median = 0.700), followed by \textit{Random–Adaptive} (mean = 0.444, SD = 0.302, median = 0.500) and \textit{Misaligned–Adaptive} (mean = 0.388, SD = 0.256, median = 0.400). Because Shapiro–Wilk tests indicated non-normality, we applied Kruskal–Wallis tests with Dunn’s post-hoc comparisons (Holm-corrected). Shown helper frequency was not detected a statistically significant effect in across conditions ($H = 2.23$, $p = 0.328$). Acceptance rate differed significantly ($H = 19.86$, $p < 0.001$), with \textit{Aligned–Adaptive} significantly higher than both \textit{Misaligned–Adaptive} ($p_{corrected} < 0.001$) and Random–Adaptive ($p_{corrected} = 0.003$). Accepted helper frequencies also showed a significant effect ($H = 8.22$, $p = 0.016$), with \textit{Aligned–Adaptive} exceeding \textit{Misaligned–Adaptive} ($p_{corrected} = 0.024$) and marginally exceeding \textit{Random–Adaptive} ($p_{corrected} = 0.048$), while no difference emerged between \textit{Misaligned–Adaptive} and \textit{Random–Adaptive}.

\section{Results}

\begin{table*}[htbp]
\centering
\caption{Results of Linear Mixed Effects Models. The Aligned-Adaptive condition was used as the reference category.}
\label{tab:lme-results}
\begin{tabular}{p{4cm}cccccc}
\toprule
 & \textbf{Coef.} & \textbf{Std. Error} & \textbf{z} & \textbf{p-value}& \textbf{CI [0.025 0.975]}  & \textbf{Sig.} \\
\midrule

\hline
\textbf{Task Performance and Effort} & & & & & & \\
\quad \textit{Task Accuracy} & & & & & & \\
\qquad Misaligned-Adaptive &-0.103  &  0.026 & -3.889 & \(<0.001\) & [-0.155, -0.051] & *** \\
\qquad Random-Adaptive & -0.070 &  0.026 & -2.637 &0.008& [-0.122, -0.018] & ** \\
\qquad Intercept & 0.617  &  0.034 & 17.988 & \(<0.001\) & [0.550, 0.684] & ***\\
\qquad Participant Variance & 0.026  &  0.089  & & & & \\
\midrule
\quad \textit{Perceived Performance} & & & & & & \\
\qquad Misaligned-Adaptive & -0.625 &   0.154 & -4.070 & \(<0.001\) & [-0.926, -0.324] & *** \\
\qquad Random-Adaptive & -0.594  &  0.154 & -3.866 & \(<0.001\) & [-0.895, -0.293] &*** \\
\qquad Intercept & 3.375  &  0.191 & 17.638 & \(<0.001\) & [3.000,  3.750] & ***\\
\qquad Participant Variance & 0.794  &  0.466  & & & & \\
\midrule
\quad \textit{Perceived Effort} & & & & & & \\
\qquad Misaligned-Adaptive & 0.562  &  0.173 & 3.256 & 0.001 & [0.224,  0.901] & ** \\
\qquad Random-Adaptive & 0.531  &  0.173 & 3.075 & 0.002 & [0.193,  0.870] & ** \\
\qquad Intercept & 2.656  &  0.190& 13.972& \(<0.001\) & [2.284,  3.029] & ***\\
\qquad Participant Variance & 0.679 &   0.377   & & & & \\
\midrule
\hline
\textbf{User Experience} & & & & & & \\
\quad \textit{Benevolence} & & & & & & \\
\qquad Misaligned-Adaptive & -0.667  &  0.174 & -3.829 & \(<0.001\) & [-1.008, -0.325] & *** \\
\qquad Random-Adaptive & -0.344   & 0.174& -1.974 &0.048& [-0.685, -0.003] & * \\
\qquad Intercept & 3.750  &  0.189 & 19.875&  \(<0.001\) & [3.380,  4.120] & ***\\
\qquad Participant Variance & 0.654  &  0.364  & & & & \\
\midrule
\quad \textit{Efficiency} & & & & & & \\
\qquad Misaligned-Adaptive & -0.383 &   0.099 &-3.871 &\(<0.001\) &[-0.577, -0.189] & *** \\
\qquad Random-Adaptive & -0.234  &  0.099 &-2.370& 0.018& [-0.428, -0.041] & * \\
\qquad Intercept & 3.906  &  0.141& 27.786& \(<0.001\) & [3.631,  4.182]& *** \\
\qquad Participant Variance & 0.476  &  0.415  & & & & \\
\midrule
\quad \textit{Dependability} & & & & & & \\
\qquad Misaligned-Adaptive &  -0.500 &   0.111 &-4.519& \(<0.001\) &[-0.717, -0.283] & ***\\
\qquad Random-Adaptive & -0.281  &  0.111& -2.542& 0.011 &[-0.498, -0.064] &* \\
\qquad Intercept & 3.602  &  0.116 &30.923& \(<0.001\) & [3.373,  3.830] & *** \\
\qquad Participant Variance & 0.238  &  0.213 & & & & \\
\bottomrule
\end{tabular}
\begin{center}
\small
\textit{Note:} Significance codes: *** p < 0.001, ** p < 0.01, * p < 0.05. All models include participant as random intercept.
\end{center}
\end{table*}

We evaluated participants’ performance, workload, and perceptions across the three experimental conditions (\textit{Aligned-Adaptive}, \textit{Misaligned-Adaptive}, and \textit{Random-Adaptive}). Our primary interest focused on pairwise comparisons of \textit{Aligned-Adaptive} vs. \textit{Misaligned-Adaptive} and \textit{Aligned-Adaptive} vs. \textit{Random-Adaptive}.

For the analysis of user-related outcomes (e.g., task accuracy, workload scores, user experience ratings), we employed LME, which provides a critical advantage for our study that, because our system is personalized and adapts to each individual’s interaction patterns, participants may differ substantially in how thresholds shift and interventions are triggered. In all models, condition was treated as a fixed effect, with the \textit{Aligned-Adaptive} condition set as the reference and participant ID as a random intercept. This allows direct interpretation of the fixed-effect coefficients for \textit{Misaligned-Adaptive} and \textit{Random-Adaptive} conditions relative to \textit{Aligned-Adaptive}. Statistical analyses were conducted in Python using the \texttt{statsmodels} and \texttt{scipy.stats} packages.

\begin{figure}[htbp]
    \centering
    \includegraphics[width=0.49\textwidth]{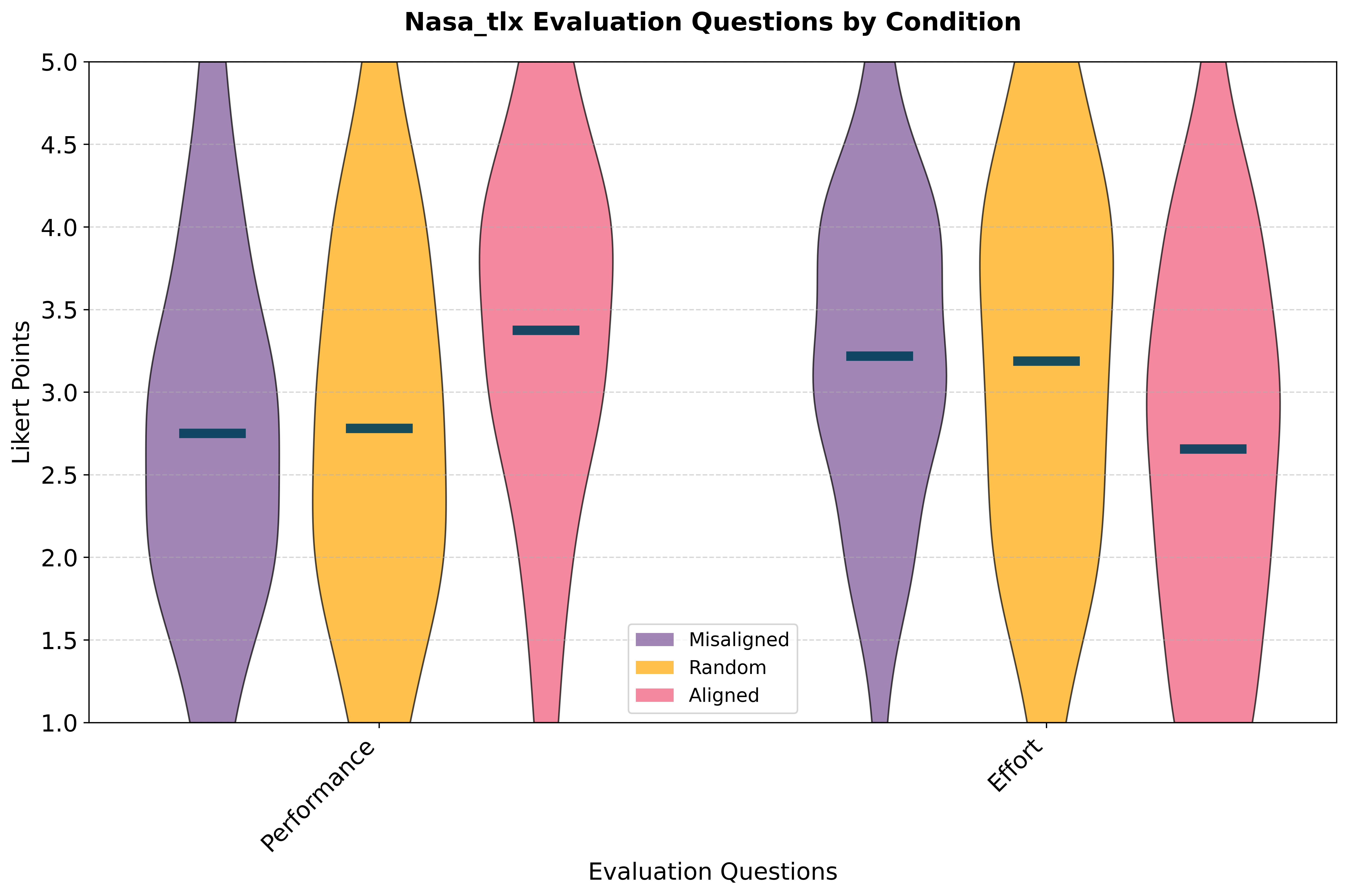} 
    \caption{Violin plot of participants’ perceived task performance and effort of the system across conditions. The distribution of responses is shown for the \textit{Aligned-Adaptive}, \textit{Misaligned-Adaptive}, and \textit{Random-Adaptive} conditions, with the slash lines indicating the mean score for each question per condition. 
    For performance, higher is better, and for effort, lower is better.}
    \Description{2 boxplots with means indicating the performance and effort dimensions for the NASA TLX results. Performance (higher is better): Means are similar for misaligned and random at around 2.75 and higher for aligned at around 3.4. Effort (lower is better): Means are similar for misaligned and random at around 3.2 and lower for aligned at around 2.6. The spread of random is wider than that of misaligned.} 
    \label{fig:tlx}
\end{figure}

\begin{figure}[htbp]
    \centering
    \includegraphics[width=0.49\textwidth]{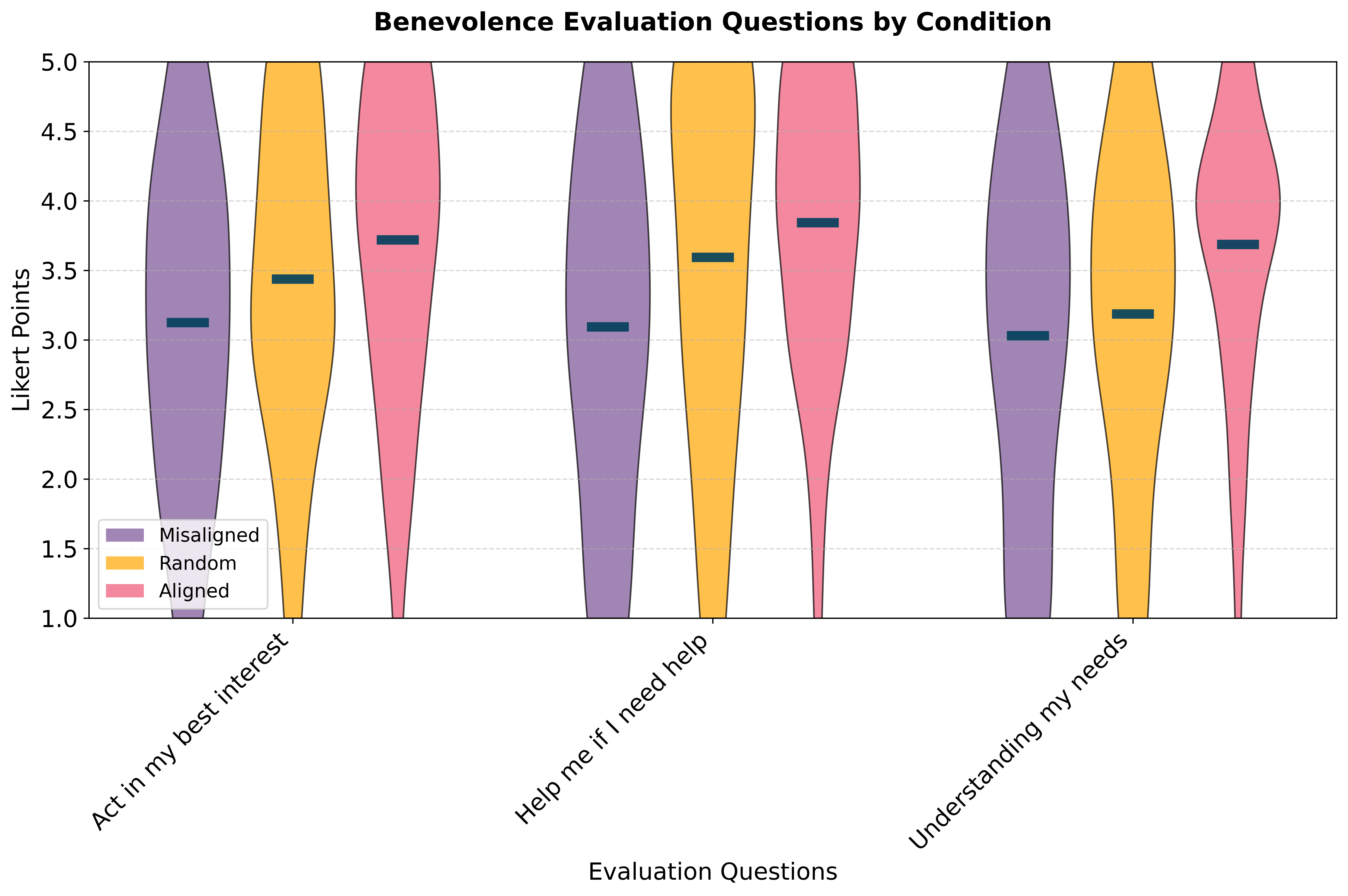} 
    \caption{Violin plot of participants’ perceived benevolence of the system across conditions. The distribution of responses is shown for the \textit{Aligned-Adaptive}, \textit{Misaligned-Adaptive}, and \textit{Random-Adaptive} conditions, with the slash lines indicating the mean score for each question per condition.}
    \Description{3 boxplots with means indicating the responses to the Likert questions "act in my best interest", "help me if I need help" and "understanding my needs". Higher is better for all questions. Act in my best interest: Means are lowest for misaligned (~3.1), followed by random (~3.4), and highest for aligned (~3.7). The spread is widest for misaligned. Help me if I need help: Means are lowest for misaligned (~3.1), followed by random (~3.6), and highest for aligned (~3.8). The spread is widest for misaligned. Understanding my needs: Means are lowest for misaligned (~3.1), followed by random (~3.2), and highest for aligned (~3.7). The spread is widest for misaligned and narrowest for aligned.} 
    \label{fig:benevolence}
\end{figure}

\subsection{Task Performance and Effort}

We analyzed participants’ task performance in terms of accuracy across the calibration block and the three experimental conditions. The mean accuracy for the calibration block was 41\% (SD = 17\%), reflecting baseline performance prior to system adaptation. Participants’ mean accuracy increased in the experimental blocks: In the \textit{Misaligned-Adaptive} condition, accuracy averaged 51\% (SD = 23\%), followed by 55\% (SD = 20\%) in the \textit{Random-Adaptive} condition. Performance was highest in the \textit{Aligned-Adaptive} condition, with a mean accuracy of 62\% (SD = 15\%). This suggests that participants performed better when interacting with the fully \textit{Aligned-Adaptive} system compared to the other conditions.   

Further, we analyzed participants’ perceived task performance and effort using item-level scores from the unweighted NASA-TLX \cite{Hart1988}, as shown in \autoref{fig:tlx} and \autoref{tab:lme-results}. For the performance dimension, where higher scores indicate better self-reported performance, the LME model revealed that participants rated their performance highest in the \textit{Aligned-Adaptive} condition. Relative to \textit{Aligned-Adaptive}, perceived performance was significantly lower in the \textit{Misaligned-Adaptive} condition and in the \textit{Random-Adaptive} condition.
For the effort dimension, where lower scores indicate lower perceived effort, the model indicated that participants reported the least effort in the \textit{Aligned-Adaptive} condition. Perceived effort increased significantly in the \textit{Misaligned-Adaptive} condition and in the \textit{Random-Adaptive} condition. 

\subsection{User Experience}
\begin{figure}[htbp]
    \centering
    \includegraphics[width=0.49\textwidth]{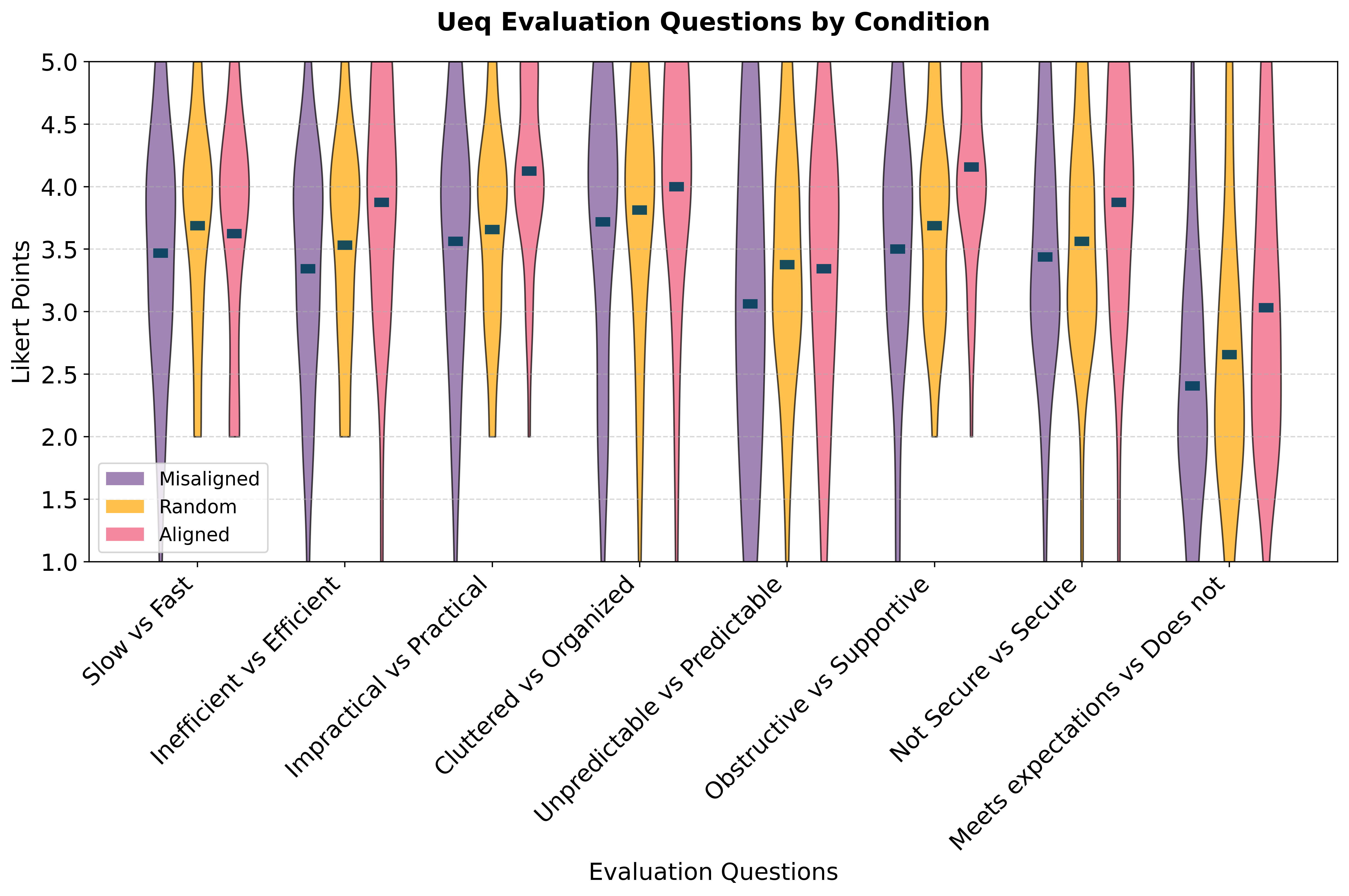} 
    \caption{Violin plot of participants’ perceived user experience of the system across conditions. The distribution of responses is shown for the \textit{Aligned-Adaptive}, \textit{Misaligned-Adaptive}, and \textit{Random-Adaptive} conditions, with the horizontal lines indicating the mean score for each question per condition.}
    \Description{8 boxplots for the UEQ adjective pairs (1) slow vs. fast, (2) inefficient vs. efficient, (3) impractical vs. practical, (4) cluttered vs. organized, (5) unpredictable vs. predictable, (6) obstructive vs. supportive, (7) not secure vs. secure, (8) meets expectations vs. does not. Higher scores are better. Means are lowest for misaligned, followed by random, and highest for aligned except for (1) and (5), where random is better than aligned.} 
    \label{fig:ueq}
\end{figure}

Participants’ perceptions of system benevolence, measured as the average of three Likert items, were generally high across all conditions, as shown in \autoref{fig:benevolence} and \autoref{tab:lme-results}. Relative to \textit{Aligned-Adaptive}, perceived benevolence decreased significantly in the \textit{Misaligned-Adaptive} condition and to a lesser extent in the \textit{Random-Adaptive}. 

As shown in \autoref{fig:ueq} and \autoref{tab:lme-results},  the average efficiency score, where higher values indicate a stronger impression that tasks can be completed effectively and without unnecessary effort, showed a significant decrease in the \textit{Misaligned-Adaptive} relative to \textit{Aligned-Adaptive} and a smaller but still significant decrease in the \textit{Random-Adaptive} condition. Similarly, for the average dependability score, reflecting the participants' sense of control over the interaction, ratings were significantly lower for \textit{Misaligned-Adaptive} and \textit{Random-Adaptive} compared to \textit{Aligned-Adaptive}.

We analyzed four individual items of particular interest, efficient, practical, supportive, and meets expectations, to capture fine-grained perceptions of the system. For the item ``efficient'', participants rated the system significantly lower in \textit{Misaligned-Adaptive} (Coef. = -0.531, SE = 0.147, z = -3.621, p < 0.001, 95\% CI [-0.819 -0.244]) and \textit{Random-Adaptive} (Coef. = -0.344, SE = 0.147, z = -2.343, p = 0.019, 95\% CI [-0.631 -0.056]) compared to \textit{Aligned-Adaptive}. Similar patterns were observed for ``practical'' (\textit{Misaligned-Adaptive}: Coef. = -0.562, SE = 0.130, z = -4.315, p < 0.001, 95\% CI [-0.818 -0.307]; \textit{Random-Adaptive}: Coef. = -0.469, SE = 0.130, z = -3.596, p < 0.001, 95\% CI [-0.724 -0.213]), ``supportive'' (\textit{Misaligned-Adaptive}: Coef. = -0.656, SE = 0.153, z = -4.277, p < 0.001, 95\% CI [-0.957 -0.356]; \textit{Random-Adaptive}: Coef. = -0.469, SE = 0.153, z = -3.055, p = 0.002, 95\% CI [-0.769 -0.168]), and ``meets expectations'' (\textit{Misaligned-Adaptive}: Coef. = -0.625, SE = 0.186, z = -3.360, p = 0.001, 95\% CI [-0.990 -0.260]; \textit{Random-Adaptive}: Coef. = -0.375, SE = 0.186, z = -2.016, p = 0.044, 95\% CI [-0.740 -0.010]). Random intercept variances ranged from 0.260 to 0.563 across items.

\subsection{Qualitative Findings}
To complement the quantitative results, we examined the transcriptions of the post-study interview data to gain deeper insights into participants’ experiences of interacting with the intelligent support agent (cf. \autoref{sec:interview}). We adopted a bottom-up thematic analysis approach, allowing patterns to emerge inductively from the data. Two researchers independently coded the transcripts, generating initial codes and grouping them into broader themes. Through iterative discussion, discrepancies were resolved, and a consensus on the final thematic results was reached, which was then systematically applied back to the full set of transcripts to ensure consistent interpretation and representation of participants’ perspectives. The resulting themes capture participants’ perspectives on the agent’s behavior, their subjective experiences with adaptive support, and broader considerations for real-world use.

\subsubsection{Motivation to Interact}
A recurring theme in the interviews was the influence of confidence and motivation on participants’ willingness to engage with the system. Many participants emphasized that their decision to accept or decline assistance was directly tied to their confidence in answering a question (N = 27). As one participant explained, “I declined only when I was really sure about the answer and otherwise whenever it popped out and I wasn't sure what's the correct answer I always went for the help” (P27). 

Interestingly, participants reported different personal thresholds of certainty before accepting help, ranging from as low as 50\% to as high as complete certainty. Some described interacting with the system even when fully confident, framing it as a form of reassurance or validation. For instance, one participant noted: “Because if I'm 100\% sure, it cost me nothing to be more sure than the 100\%. But sometimes the human brain says, I'm sure, I don't need help. But at the end, I think the AI help is much better. Or with the AI help, it's much better. No matter how sure you are, it's better to be more sure” (P5). The agent functioned not only as a compensatory tool in moments of uncertainty but also as a confidence amplifier.

Motivational factors also played a substantial role. Participants indicated that a desire to perform well or achieve higher accuracy influenced their likelihood of accepting assistance, especially when they perceived the task as evaluative or high-stakes (N = 18). Beyond performance, curiosity emerged as an additional motivator (N = 7). Participants explicitly described engaging with the agent out of an interest in learning new information, with one participant explaining: “That motivated me to learn more new things and get more new information” (P2).

These findings highlight that the decision to interact with the agent was shaped by a combination of situational confidence, performance goals, and intrinsic curiosity.

\subsubsection{Helpfulness and Interruptions}
Participants generally did not perceive the agent as disruptive to their workflow (N = 30). Participants explicitly described the system as unobtrusive and likened its presence to a supportive companion rather than an interruption. As one participant reflected, “It didn’t interrupt my workflow at all. It felt like when someone is thirsty and someone is like, do you need water or something, it was really helpful” (P4). This characterization highlights how the agent was largely experienced as a contextual aid, intervening only when relevant without negatively affecting task engagement.

Nonetheless, some limitations in the presentation of assistance were noted. Some participants reported frustration when the system’s responses were slightly verbose or required additional effort to extract the information most relevant to the task (N = 7). For these participants, the need to parse lengthy or imprecise responses reduced the immediacy of the support. At the same time, preferences varied: others valued more elaborate explanations that provided additional context or opportunities for incidental learning (N = 3). As one participant noted, “I think in a way they were still pretty helpful where it still gives you some sort of information. Maybe not the specific thing. You still learn something exactly” (P11).

These findings suggest that participants widely perceived the agent as helpful and minimally disruptive, but expectations about the granularity of responses differed. This points to the importance of personalization in balancing concise task support with opportunities for broader learning.

\subsubsection{Usage and Applications}
Across conditions, participants consistently described differences in how and when support was offered, underscoring that the timing of adaptive interventions shaped their experience as much as the content of the support itself. They explicitly noted perceivable differences between conditions, often emphasizing mismatches between when they expected assistance and when it actually appeared (N = 16). For example, some described that help did not align with the moments they most needed it: ``Across different blocks and among those questions, not always the AI gave me the help that I needed'' (P3) and ``For my first block, it appeared later in questions. For the second one, it appeared kind of randomly'' (P14). Others contrasted two conditions based purely on temporal predictability: ``In the first block, it was confused and with few help, but in the second one, I was chilling because I knew the AI would help me anyway. The first one was more stressful'' (P5). Multiple participants described heightened cognitive load and negative affect when interventions arrived too late or not at all, creating unfulfilled expectations: ``In the first block, the system helped me more. Then in the other block it didn't help sometimes, so I kept waiting for it, but it didn't showed up'' (P13). This waiting behavior indicates that participants developed temporal expectations based on past experience, and violations of these expectations, even when help eventually arrived, created frustration: ``It made me frustrated because I was waiting for help but didn't get it'' (P24). These observations highlight that the perceived effectiveness of cognition-aware support hinges not only on what help is provided, but critically on when it is delivered.

Furthermore, participants frequently discussed how such a system could extend beyond the experimental setting into everyday contexts. Education, healthcare, and other professional processes were mentioned as particularly promising application areas, especially in situations where complex terminology or specialized language presents barriers to understanding. For instance, one participant reflected on healthcare contexts: “Maybe like a survey from doctors or hospitals [...] when filling those paper surveys I do feel I use Google myself. There’s some complex terminology, too scientific. A system like this would help” (P32). Similarly, others highlighted bureaucratic contexts such as banking or government forms: “I could see it actually helping me with filling forms in bureaucracy because I have to google a lot of things and having a system like this would offer some help” (P27).

Finally, participants expressed preferences for different levels of control over how and when help should be triggered. While some (N = 22) valued the system’s proactive interventions, others (N = 3) preferred greater autonomy, suggesting that “always-on” availability or customizable triggers would allow them to adapt the system to their own working styles, which highlights the importance of flexible design to accommodate diverse expectations.

\section{Discussion}

We present a personalized, rule-based adaptive support system to mitigate cognitive overload in knowledge-intensive online tasks and examine how the timing of AI assistance affects user experience. Our primary methodological contribution is a real-time overload detection mechanism that integrates multimodal signals, EDA and mouse movement, using individually calibrated, continuously updated thresholds. A second contribution is a systematic comparison of three adaptation strategies: \textit{Aligned-Adaptive} (assistance timed to inferred cognitive states), \textit{Misaligned-Adaptive} (deliberately mistimed), and \textit{Random-Adaptive} (arbitrary timing), to assess the impact of timing in proactive support systems.

\subsection{Summary of Results}

Our adaptive system improved both performance and user experience. Task accuracy rose from 41\% at baseline to 62\% in the \textit{Aligned-Adaptive} condition, compared with 51\% and 55\% in the other conditions. Participants also reported higher perceived performance and lower mental effort with aligned assistance.

Trust and user experience benefited as well: efficiency, dependability, and benevolence ratings were significantly higher in the \textit{Aligned-Adaptive} condition, consistent with human–technology trust frameworks \cite{https://doi.org/10.48550/arxiv.2507.21158, Duan2024}.

Qualitative analysis showed that timely interventions were perceived as supportive, whereas misaligned timing caused stress and frustration. Help-seeking behavior was influenced by confidence, motivation, and curiosity. Participants envisioned applications in healthcare, education, and professional workflows.

\subsection{Adaptive Support Improves Performance}

We expected that response accuracy would be significantly higher in the \textit{Aligned-Adaptive} system than in the \textit{Misaligned-Adaptive} or \textit{Random-Adaptive} systems.
The results confirmed this expectation: participants achieved the highest accuracy in the \textit{Aligned-Adaptive} condition, outperforming both other conditions. These findings echo evidence from survey methodology that comprehension aids can reduce errors \cite{Conrad2000, Conrad2007}. Previous research also shows that respondents rarely request help on their own: Often fewer than 10\% use optional clarifications \cite{conrad2006use, conrad2003interactive}, highlighting the value of systems that can proactively offer assistance when needed. Our results extend this literature by demonstrating that the effectiveness of proactive support depends on its alignment with the respondent’s cognitive state. Participants in the \textit{Misaligned-Adaptive} and \textit{Random-Adaptive} conditions did not experience the same improvements, suggesting that when assistance is offered matters. In our study, the \textit{Aligned-Adaptive} condition did not trigger substantially more often overall, yet it produced a significantly higher acceptance rate, indicating that support perceived as timely or appropriate is more likely to be taken up, and this higher uptake is what ultimately contributes to improved response quality. Our findings should not be interpreted as evidence about static, passively available personalization models (i.e., non-adaptive), since these factors were not experimentally manipulated. Instead, the comparison among aligned, misaligned, and random triggering suggests a design implication grounded directly in our data: adaptive assistance is most beneficial when its triggering logic produces support at moments that correspond to respondents’ actual task demands. Rather than emphasizing classifier precision abstractly, our results indicate that timing that feels meaningfully connected to respondents’ present experience, is essential for improving response quality.

\subsection{Adaptive Systems Enhance User Experience}

We expected that the \textit{Aligned-Adaptive} system would lead to superior perceived user experience, reflected in higher ratings of efficiency, dependability, and benevolence. This hypothesis was also supported. Participants rated the \textit{Aligned-Adaptive} system more efficient, dependable, and benevolent than the \textit{Misaligned-Adaptive} and \textit{Random-Adaptive} controls. These qualities are central to user trust: benevolence and dependability are established antecedents of system acceptance in HCI and organizational psychology \cite{Mayer1995, McKnight2002}.

These findings are grounded in established results in adaptive assistance research, showing that help perceived as proactive but non-intrusive increases user satisfaction \cite{Andolina2018}. They also connect to recent advances in proactive LLM systems across diverse domains. In programming, adaptive LLM-based support has been shown to boost developer efficiency when aligned with task context \cite{Chen2025,Pu2025}; in collaborative and social settings, proactive interventions can strengthen perceptions of support and benevolence \cite{Yang2025,Liu2024}. However, prior work has also warned that mistimed interventions risk undermining trust by appearing disruptive \cite{Prasongpongchai2025}. Our results empirically substantiate this: when assistance was mistimed (\textit{Misaligned-Adaptive} condition), participants rated the system as less efficient, dependable and benevolent.

We show that adaptive timing not only preserves data quality but also fosters perceptions of efficiency, dependability, and benevolence, introducing a trust-oriented lens.

\subsection{Personalized Support Reduces Workload}
We expected that participants would report lower subjective workload when interacting with the \textit{Aligned-Adaptive} system compared to the \textit{Misaligned-Adaptive} or \textit{Random-Adaptive} systems. 
This hypothesis was confirmed: subjective workload ratings (NASA-TLX effort) were lowest and perceived performance highest in the \textit{Aligned-Adaptive} condition. A key factor underlying this effect appears to be participants’ greater willingness to accept assistance in the aligned condition. Although the \textit{Aligned-Adaptive} system did not deliver substantially more interventions overall, it yielded a significantly higher acceptance rate, resulting in a larger number of help episodes actually taken up. This suggests that support perceived as relevant or well-timed is more likely to be used, and that this higher uptake, rather than sheer availability, contributes to reductions in perceived workload. Participants may have felt more supported not because the system intervened more frequently, but because its interventions were judged as meaningful enough to accept.

Task-duration metadata provide additional context for these effects. Although accepting help can introduce extra interaction steps, these differences in task duration were not statistically significant. The \textit{Aligned-Adaptive} condition, despite a higher rate of accepted interventions, did not result in longer task times compared to the \textit{Misaligned-Adaptive} condition. This pattern is consistent with the idea that well-timed, relevant support can reduce cognitive overload, allowing users to keep efficiency even engaging with assistance.

These results complement prior findings that proactive assistance can reduce effort by easing search, interpretation, and comprehension demands \cite{Andolina2018}. They also extend survey research showing that cognitive strain contributes to satisficing, careless responding, and breakoff \cite{Peytchev2009,Liu2017}. Our study advances this literature by demonstrating that multimodal sensing, EDA, a well-established indicator of sympathetic arousal and workload \cite{Boucsein2012,Kosch2023}, together with mouse dynamics \cite{horwitz2017using}, can be operationalized in real time to deliver support that participants are more willing to accept.

Rather than claiming that physiological and behavioral responsiveness alone “maintains” cognitive capacity, our findings more cautiously point to a design implication: systems that adapt to users’ moment-to-moment experience in ways they recognize as appropriate can meaningfully reduce perceived workload, especially in extended, sequential tasks where cumulative effort matters.

\subsection{Over-Reliance, Reassurance-Seeking, and Shifts in User Agency}

Our findings surface important ethical considerations for cognition-aware support systems, particularly the risk of over-reliance and reassurance-seeking behaviors. Several participants described anticipating or waiting for the system’s help, and some expressed frustration when support did not appear as expected. These accounts suggest that users may begin to defer judgment or delay action in hopes of being assisted, indicating a subtle shift in agency from user to system. Such patterns resonate with recent evidence that users’ self-confidence and perceived competence strongly shape their reliance on AI: increases in trust or decreases in confidence lead to greater relative AI reliance \cite{Schemmer2023}. In our study, the higher acceptance rate in the aligned condition may similarly reflect a reinforcing loop in which timely support boosts trust, which in turn may heighten reliance on the system’s guidance. While this can improve task performance, it risks creating dependencies that undermine long-term autonomy or diminish users’ confidence in their own abilities. 

At the same time, participants voiced divergent preferences regarding control over how and when assistance should be triggered. A substantial majority (N = 22) explicitly favored proactive, system-initiated interventions, describing them as convenient, reassuring, or cognitively easing. While this preference speaks to the perceived value of timely support, it also signals an emergent ethical risk: when users overwhelmingly prefer to be acted upon rather than to act, cognition-aware systems may drift toward increasingly interventionist defaults, gradually shifting normative expectations around when help should appear and who, human or system, directs the flow of interaction. Such asymmetries risk reinforcing over-reliance by making proactive intervention feel natural and self-directed effort comparatively burdensome. Others (N = 3) explicitly preferred greater autonomy and suggested that “always-on” availability or customizable triggering options would better align with their personal working styles. These contrasting preferences underscore the ethical importance of designing adaptive systems that are not only responsive but also flexible, allowing users to calibrate the degree of automation and maintain meaningful control over their interaction. Collectively, these findings highlight a broader ethical imperative: cognition-aware systems must balance helpfulness with transparency and user agency, ensuring that adaptive support augments rather than eclipses human decision-making. Long-term deployments should examine whether and how repeated exposure amplifies reliance patterns, reshapes expectations of assistance, or alters users’ sense of competence and control.

\subsection{Design Implications}

Our study reveals core insights for designing adaptive AI support systems in cognitively demanding environments. 

\subsubsection{Multimodal sensing is feasible and practical} The technical foundation demonstrates that multimodal sensing can work in practice: combining EDA and mouse movement provides sufficient signal for real-time cognitive load detection. This approach offers a practical alternative to more invasive methods like EEG, enabling deployment in everyday digital environments where users cannot be expected to wear specialized equipment.

\subsubsection{Prioritize Temporal Precision Over Content Sophistication} While much HCI research focuses on improving the quality of help content (e.g., more detailed explanations \cite{Ma2025}, better visualizations \cite{Mackinnon2025}, multimodal presentations \cite{https://doi.org/10.12758/mda.2025.09}), our results suggest that when help arrives may matter in addition to what is delivered. Systems should invest in accurate prediction of moments of user difficulty rather than solely optimizing content. A simple, well-timed explanation delivered at the moment of confusion may be more effective than a sophisticated explanation that arrives too early (causing interruption) or too late (after the user has already made errors or moved on). This has practical implications for proactive system design: intelligent collaborations may achieve greater impact by improving temporal detection mechanisms (e.g., physiological sensors, behavioral analytics, machine learning models) than by iterating on help content alone.

\subsubsection{Account for Individual Differences in Struggle Signals} Our personalized classifiers with dynamic threshold adaptation proved critical. What signals struggle for one user (e.g., 15 seconds of mouse hesitation) may represent normal processing speed for another. This finding extends beyond surveys to any domain where users vary in baseline behavior, expertise, or cognitive processing styles. Adaptive systems should calibrate to individual users through brief onboarding tasks or continuous background monitoring rather than applying population-level heuristics, and implement personalized baselines that update throughout the session as the system learns each user's typical patterns. For example, educational software should distinguish between a typically fast student pausing versus a typically slow student taking the same time.

\subsubsection{Design for Temporal Expectation Management} Participants developed expectations about when help would arrive based on prior interaction patterns, and violations of these expectations caused frustration even when help quality remained constant. Consistency in intervention timing may be as important as accuracy. Users appear to build mental models of system behavior that influence their cognitive strategies (e.g., whether to persist through difficulty or wait for assistance). Systems should either maintain consistent temporal patterns or explicitly communicate when they are operating in different modes. If temporal patterns vary, provide subtle indicators of system state. For instance, a decision-support tool might signal ``monitoring mode'' versus ``active assistance mode'' so users know whether to expect immediate help.

\subsubsection{Mitigate Cascading Effects of Missed Interventions} Our results demonstrated spillover effects where early struggles compounded across subsequent items. This highlights that the cost of false negatives (missing a moment of struggle) may exceed the cost of false positives (offering unnecessary help) in sequential task contexts. A single missed intervention can trigger satisficing behaviors, reduced engagement, and degraded performance on later items, creating cumulative harm that exceeds the momentary annoyance of occasionally receiving unnecessary help. Therefore, in domains with sequential dependencies (educational assessments, multi-step procedures, complex forms), calibrate interventions to favor recall over precision, better to occasionally offer help when not needed than to miss critical moments of struggle. Implement ``recovery mechanisms'' that detect when users have fallen into suboptimal patterns (e.g., rapid clicking, random responses) and proactively offer support to break the cascade before it progresses further.

\subsubsection{Distinguish Between Productive Struggle and Unproductive Floundering} Not all difficulty requires intervention. Educational research distinguishes between productive struggle (moderate challenge that promotes learning) and unproductive floundering (excessive difficulty that causes frustration without benefit) \cite{Sinha2021}. The adaptive system should navigate this distinction, intervening when users flounder but avoiding premature help that short-circuits beneficial struggle. Implement multi-stage intervention strategies that begin with minimal support (e.g., subtle hints, progress indicators) and escalate only if struggle persists beyond a productive threshold. For example, an intelligent tutoring system might first highlight relevant materials, then provide a conceptual hint, and finally offer worked examples. This graduated approach respects productive struggle while preventing unproductive floundering. Physiological signals may help distinguish these states: moderate EDA increases suggest productive engagement, while sharp spikes or prolonged elevated levels indicate unproductive stress.

\subsubsection{Balance Proactivity with User Agency} While proactive intervention improved outcomes, some participants reported wanting more control over when they received help. Overly aggressive support can undermine user autonomy and create learned helplessness where users stop attempting problems independently. Systems must balance predictive proactivity with user agency. Provide configurable intervention modes that users can adjust based on personal preferences and task context. Options might include: (1) ``Proactive'' mode where the system intervenes automatically based on predicted struggle, (2) ``Suggestive'' mode where the system signals help availability but requires explicit acceptance, (3) ``On-demand'' mode where users must request help manually. Allow users to switch modes mid-task as their confidence or task difficulty changes. Track mode preferences and performance to learn optimal defaults for different user profiles.

\subsubsection{Design for Transparency in Timing Decisions} Participants sometimes expressed confusion about why help appeared when it did (or didn't). Consider incorporating lightweight explanations of timing logic, particularly when users express confusion. For example: ``I noticed you paused on this question, would you like help?'' or ``You've been working on similar problems successfully, so I'm giving you space to try this independently.'' This transparency can help users develop more accurate mental models of system behavior and reduce frustration from unexpected timing. However, explanations must be brief to avoid adding cognitive load during already-challenging moments.

\subsubsection{Generalization Across Domains} While our study focused on surveys with knowledge-based questions, the principle of temporal alignment applies broadly to any interactive system where users encounter intermittent cognitive demands. 

Temporal precision of hint delivery may be as consequential as hint content quality. Rather than offering support on fixed schedules or after predetermined error counts, systems could leverage behavioral and physiological indicators to detect when individual students transition from productive struggle (which promotes learning) to unproductive floundering (which causes disengagement), triggering interventions precisely at this inflection point.

Medical professionals may encounter diagnostic uncertainty, yet existing systems typically present information reactively in response to explicit queries. Systems could monitor interaction patterns, dwell times, and consultation sequences to identify moments when clinicians exhibit uncertainty, proactively surfacing relevant evidence or differential diagnoses at these critical junctures rather than waiting for manual information retrieval.

Traditional approaches to software assistance rely on upfront tutorials or context-sensitive help menus, which users often ignore or find intrusive. Temporally aligned systems could instead detect when users attempt tasks inefficiently (e.g., using multiple manual steps for operations that could be automated) and deliver just-in-time feature suggestions at the moment of inefficiency, thereby reducing interruption costs while maintaining relevance.

In collaborative platforms, some participants may struggle to contribute effectively due to domain knowledge gaps, language barriers, or social anxiety. Rather than providing uniform scaffolding to all participants, temporally aligned systems could detect when specific individuals encounter barriers through participations' interaction patterns, communication hesitancy, or physiological signals, offering targeted support (e.g., background information, prompts, or facilitation) precisely when engagement falters.

These applications share in common: systems must continuously sense user state, predict moments of cognitive demand or uncertainty, and trigger interventions with temporal precision. The specific sensing modalities will necessarily vary by domain, physiological signals and mouse dynamics for individual computer tasks, interaction logs and performance patterns for software applications, communication patterns and contribution metrics for collaborative systems, but the fundamental design imperative remains constant.

\subsection{Limitations \& Future Work}

Our study demonstrates the potential of adaptive, data-driven support systems, but several limitations remain. 

The controlled laboratory setting and short task duration limit the ecological validity of these findings. The demographic composition of our sample was biased toward younger and middle-aged adults, potentially limiting the generalizability of our findings to older populations, who may exhibit different cognitive and physiological patterns. In real-world surveys, interactions often span longer periods, occur across diverse devices, and take place under multitasking conditions, all of which may affect cognitive load and help-seeking behavior. Future research should extend this work to field deployments on online platforms and mobile devices to better assess generalizability. However, the modalities we analyze are increasingly accessible outside the lab: behavioral signals such as mouse movement and response timing are already available in virtually all online survey platforms \cite{horwitz2017using, horwitz2020learning}; physiological indicators such as EDA can be captured through common consumer wearables \cite{Lazarou2024, Siirtola2019}. Major survey panels and commercial platforms have begun piloting integrations with smartwatch data, suggesting a realistic pathway for deploying adaptive mechanisms at scale \cite{Kapteyn2024}. We therefore see our findings not as limited to laboratory use but as a foundation for future real-world studies that exploit emerging, low-burden sensing technologies. Evaluating how these adaptive systems perform under naturalistic noise conditions remains an important next step.

Our evaluation also focused on knowledge-based multiple-choice questions and a single form of intervention through textual, LLM-based assistance. However, our core sensing modalities, EDA and mouse movement, capture domain-general indicators of cognitive load and behavioral uncertainty that should transfer across item types. Open-ended responses may require further adjustment to distinguish composition effort from struggle. Grid questions may produce different mouse patterns due to spatial navigation demands. Attitudinal items may generate distinct patterns reflecting evaluative rather than retrieval processes. Our personalized, continuously adapting classifiers were designed to accommodate such variations by calibrating to individual baselines, but the extent of generalization across substantially different cognitive processes remains an empirical question. Further, the adaptive rules in our system were optimized around task accuracy, a meaningful and measurable signal for knowledge-based tasks. Other relevant outcome metric will suitable to apply. In self-report settings, indicators such as satisficing behaviors (e.g., skipping or straightlining), response consistency, or attention-check performance often provide more direct measures of data quality. Future work could therefore redesign adaptive triggers around these markers to better align with other methodological goals of survey research. At the same time, our findings offer a proof of concept: even when tuned to accuracy, adaptive timing meaningfully shaped user experience and behavior, suggesting that similar mechanisms could be repurposed to target survey-specific quality indicators once those metrics are integrated into the adaptive logic.

A central challenge lies in personalization. The system relied on a brief calibration phase to establish individual thresholds, which highlighted the difficulty in adaptive systems. The main difficulty was a cold-start problem (N = 6): when only limited calibration data were available, the system occasionally set overly sensitive thresholds, resulting in an increased number of assistance triggers early in the task and a temporary reduction in the contrast between conditions. This effect was short-lived: thresholds stabilized quickly once the rule-based updates had incorporated several observations, after which these participants experienced no abnormal triggering behavior. While such cold-start sensitivity is a common issue in personalized adaptive systems, its practical impact in real-world survey deployments may be minimal. Many surveys already begin with warm-up items (demographic questions such as age or education), which could naturally supply the initial behavioral data needed for calibration before cognitively demanding items appear. Future work could further mitigate cold-start effects by leveraging these early low-stakes questions. Technically, one promising direction is context-aware adaptation, where additional information, such as time-on-task, behavioral variability, or short-term performance fluctuations, are used to refine early load estimates and prevent premature triggering. Another improvement involves mid-task manipulation checks, in which self-report prompts are inserted at strategic points to verify whether personalized thresholds remain aligned with participants’ subjective experience. Confidence-weighted triggering could reduce early false positives by estimating predictive uncertainty (through variance measures or bootstrap ensembles) and temporarily down-weighting or gating intervention decisions when model confidence is low. These approaches would make estimation more robust, reduce sensitivity to noisy initial data, and support more stable adaptation throughout the interaction. 

We used a non–task-pretrained LLM for generating clarifications. Although we constrained outputs through carefully designed prompts and a 160-token limit, some participants still perceived the explanations as overly verbose. Future systems could refine LLM outputs through domain-specific pretraining, style tuning, or adaptive summarization mechanisms that tailor the length and focus of clarifications to both the task and the user’s interaction history. Beyond survey contexts, such refinements point to promising applications in formative educational settings, where the goal is not summative evaluation but scaffolding learners’ reasoning processes. In these environments, an LLM operating in the background could be prompted to behave more like a reflective tutor, offering hints, asking metacognitive questions, or helping learners articulate their understanding without revealing answers outright.

Further, about participant expectations, knowing the AI might provide help could have influenced subjective ratings. In misaligned conditions, participants may have perceived the system as malfunctioning, potentially exaggerating differences in dependability ratings, consistent with expectancy violation theory \cite{Burgoon2015}. However, participants were unaware of which condition they were in and only knew the system might generally use cognitive load detection. This general framing should have affected all conditions approximately equally, preserving the validity of relative comparisons between conditions. Future work could employ between-subjects designs to further explore the effects from expectancy effects.

\section{Conclusion}

This study investigated adaptive, personalized LLM-based agents for mitigating cognitive overload in web-based tasks. In a within-subject experiment with 32 participants, we found that adaptive timing improved in acceptance rate, accuracy, workload, and perceptions of efficiency, dependability, and benevolence. These results show that well-timed, personalized support not only prevents performance loss but also fosters trust and a positive experience.

Our work advances adaptive support systems by demonstrating that closed-loop interventions driven by multimodal physiological and behavioral signals can effectively predict and respond to individual cognitive struggle in real time. This temporal design focus that intervention timing matters has implications across interactive systems where users encounter variable cognitive demands. Future research should extend this approach to longer-term deployments in other survey contexts, explore generalization across diverse item formats (open-ended responses, attitudinal scales, grid questions), and investigate alternative outcome metrics aligned with other survey methodology goals. More broadly, this work points toward a new generation of adaptive agents that continuously sense user cognitive state through physiological and behavioral signals, predict moments of struggle with personalized models, and provide proactive support aligned to individual needs. By demonstrating that such systems can maintain data quality, improve task performance, and earn user trust, we establish a foundation for physiology-aware interactive systems that offer timely, contextually appropriate assistance across domains.

\section{Open Science \& Transparency}
We encourage readers to reproduce and extend our results. Therefore, our system is attached with the supplementary materials. During the preparation of this work, the authors used OpenAI’s GPT-4o and Grammarly for grammar and style editing. All content was reviewed and edited by the authors, who take full responsibility for the final publication.
%
%

\begin{acks}
This research was supported by the Federal Ministry of Research, Technology and Space (BMFTR) under grant agreement 16DKZ2019B (KODAQS). It was co-funded by the European Union – NextGenerationEU, the Deutsche Forschungsgemeinschaft (DFG, German Research Foundation), project numbers 529719707; 396057129, and the Munich Center for Machine Learning (MCML).
\end{acks}

\bibliographystyle{ACM-Reference-Format}
\bibliography{bibliography}

\clearpage
\appendix
\section{Feature Importance Analysis using F-regression}

\label{sec:featureimp}
\begin{figure}[htbp]
\centering
\includegraphics[width=0.50\textwidth]{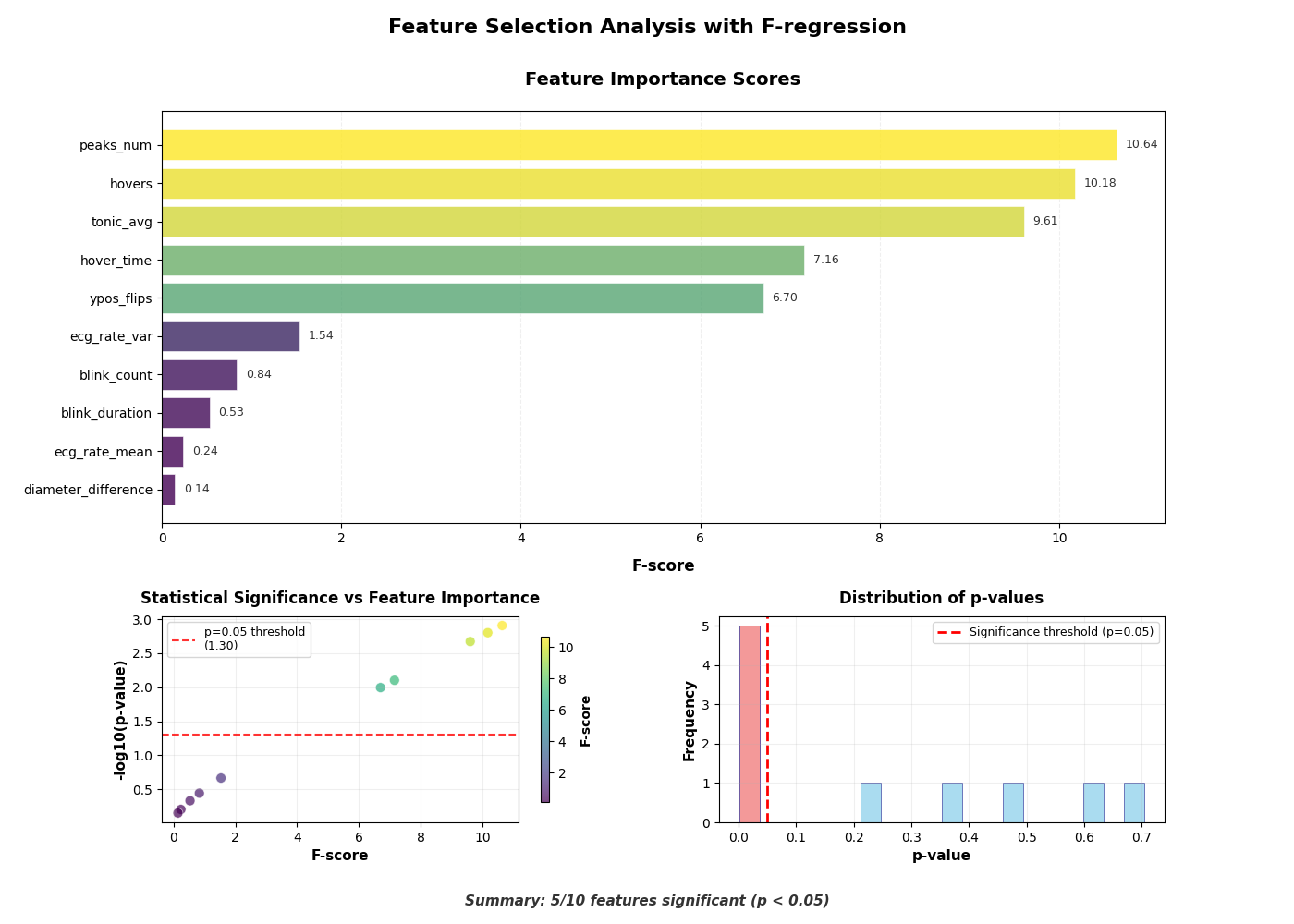}
\caption{Visualization of feature importance scores derived from F-regression analysis. The horizontal bar chart (top) displays the influential features ranked by their F-scores, with color intensity indicating relative importance. The scatter plot (bottom left) illustrates the relationship between feature importance and statistical significance (-log10(p-value)), with the red dashed line indicating the p=0.05 significance threshold. The histogram (bottom right) shows the distribution of p-values across all features, highlighting the proportion of statistically significant predictors.}
\Description{Top: Bar chart indicating feature importance scores. peaks num: 10.64, hovers: 10.18, tonic avg: 9.61, hover time: 7.16, ypos flips: 6.70, ecg rate var: 1.54, blink count: 0.84, blink duration: 0.53, ecg rate mean: 0.24, diameter difference: 0.14
Bottom left: statistical significance vs. feature importance. X-axis: F-score, y-axis: -log10(p-value). The features peaks num, hovers, tonic avg., hover time, and ypos flips are above the negative log-scaled significance threshold, all others below.
Bottom right: distribution of p-values: 5 p-values are below 0.05, the 6 others at 0.2 or higher.}
\end{figure}

\autoref{sec:featureimp} visualized the feature selection considerations.

\section{A Prompt Used in Clarification Generation}
\label{sec:prompt}
\textit{<<SYS>>\\
    You are a helpful assistant who clearly explains concepts in English. Provide ONLY the context.\\
    <</SYS>>\\
    Explain the concept "\{words\}" in English.}\\

\section{The Post-study Semi-structured Interview Questions}
\label{sec:interview}
\subsection{Overall Experience:}
\begin{itemize}
\item You just used a system that attempted to predict how difficult a question was for you and provided help based on that prediction. Can you describe your overall experience with it, especially how it felt across different questions or sessions?
\end{itemize}

\subsection{Perceived Costs and Risks:}
\begin{itemize}
\item Did accepting the help ever interrupt your workflow or make the task more difficult in any way? (For instance, did it feel mistimed, break your focus, or require extra effort to understand?)
\item What about when you declined help? Did that ever make things harder later on? (Like missing out on useful suggestions or feeling unsure afterward?)
\end{itemize}

\subsection{Decision to Accept or Decline Help:}
\begin{itemize}
\item During the experiment, we ran three different blocks of the task: one where the system adapted positively based on your behavior, one where it adapted in the opposite direction, and one where behavior was responded to randomly. Did you notice any differences between these three blocks? How so?
\item Can you walk me through your experience in any block, like what felt helpful, what felt confusing, or annoying?
\item What motivated you to accept, ignore, or reject the help when it was offered?
\item Were there times when you declined help even though you weren’t confident, or accepted help even when you felt you didn’t need it?
\item What influenced that decision?
\end{itemize}
 
\subsection{Real-World Use \& Improvements:}
\begin{itemize}
\item If you could improve the system, what would you change? (Anything that frustrated you, or features you wish it had?)
\item Would you want to use a system like this in real life? Why or why not? (And in what kind of situations or tasks would it make sense or not?) 
\end{itemize}

%
%
\end{document}